\documentclass[useAMS,usenatbib]{mn2e}

\usepackage{lscape}
\usepackage{epsfig}
\usepackage{longtable}
\usepackage{ltxtable}
\usepackage{array}

\title[Optical properties of high-frequency radio sources]{Optical Properties of High-Frequency Radio Sources from the Australia Telescope 20\,GHz (AT20G) Survey}
\author[Mahony et al.]{Elizabeth K. Mahony$^{1,2}$\thanks{E-mail:emahony@physics.usyd.edu.au}, Elaine M. Sadler$^{1}$, Scott M. Croom$^{1}$, Ronald D. Ekers$^{2}$, \and Keith W. Bannister$^{1,2}$, Rajan Chhetri$^{2,3}$, Paul J. Hancock$^{1}$, Helen M. Johnston$^{1}$, \and  Marcella Massardi$^{4,5}$ and Tara Murphy$^{1,6}$\\
$^{1}$Sydney Institute for Astronomy, School of Physics, The University of Sydney, NSW 2006, Australia\\
$^{2}$Australia Telescope National Facility, CSIRO Astronomy and Space Science, P.O Box 76, Epping, NSW 1710, Australia\\
$^{3}$Department of Astrophysics and Optics, School of Physics, The University of New South Wales, NSW 2052, Australia \\
$^{4}$Istituto di Radioastronomia, INAF, Via Gobetti 101, 40129  Bologna, Italy \\
$^{5}$INAF - Osservatorio Astronomico di Padova, Vicolo dell'Osservatorio 5, I-35122 Padova, Italy \\
$^{6}$School of Information Technologies, University of Sydney, NSW 2006, Australia 
}
\begin{document}

\date{Accepted 2011 .... Received 2011 ...; in original form 2011 ...}

\pagerange{\pageref{firstpage}--\pageref{lastpage}} \pubyear{2011}

\maketitle

\label{firstpage}

\begin{abstract}

Our current understanding of radio-loud AGN comes predominantly from studies at frequencies of 5\,GHz and below. With the recent completion of the Australia Telescope 20\,GHz (AT20G) survey, we can now gain insight into the high-frequency radio properties of AGN. This paper presents supplementary information on the AT20G sources in the form of optical counterparts and redshifts. Optical counterparts were identified using the SuperCOSMOS database and redshifts were found from either the 6dF Galaxy survey or the literature. We also report 144 new redshifts. For AT20G sources outside the Galactic plane, 78.5\% have optical identifications and 30.9\% have redshift information. The optical identification rate also increases with increasing flux density. Targets which had optical spectra available were examined to obtain a spectral classification.

There appear to be two distinct AT20G populations; the high luminosity quasars that are generally associated with point-source optical counterparts and exhibit strong emission lines in the optical spectrum, and the lower luminosity radio galaxies that are generally associated with passive galaxies in both the optical images and spectroscopic properties. It is suggested that these different populations can be associated with different accretion modes (cold-mode or hot-mode). We find that the cold-mode sources have a steeper spectral index and produce more luminous radio lobes, but generally reside in smaller host galaxies than their hot-mode counterparts. This can be attributed to the fact that they are accreting material more efficiently. Lastly, we compare the AT20G survey with the S-cubed semi-empirical (S3-SEX) models and conclude that the S3-SEX models need refining to correctly model the compact cores of AGN. The AT20G survey provides the ideal sample to do this.
\end{abstract}


\section{Introduction} \label{intro}

The formation and evolution of AGN, and the role they play in galaxy evolution, is currently one of the most active fields in astronomy. The advent of large wide-area surveys across a multitude of frequencies, from radio to gamma--rays, has provided large, homogenous samples allowing us to study the different AGN populations in a statistically significant manner. 

The radio sky at frequencies of 5\,GHz and below has been well explored, largely due to surveys such as the 843\,MHz Sydney University Molonglo Sky Survey (SUMSS; \citealt{sumss}), 1.4\,GHz NRAO VLA Sky Survey (NVSS; \citealt{nvss}) and the 5\,GHz Parkes-MIT-NRAO (PMN; \citealt{pmn}) and Green Bank 6\,cm (GB6; \citealt{gb6}) surveys. The Australia Telescope 20\,GHz (AT20G) Survey provides an unprecedented view of the high-frequency radio sky. This blind survey was carried out from 2004--2008 on the Australia Telescope Compact Array (ATCA) and covers the entire southern sky excluding $|b|<1.5^{\circ}$. The AT20G catalogue consists of 5890 sources above 40\,mJy \citep{at20g}. To obtain reliable spectral index information, observations at 4.8 and 8.6\,GHz were carried out within 1 month of the 20\,GHz observations for most sources south of $\delta=-15^{\circ}$.

Selecting sources at 20\,GHz provides a unique sample that is dominated by active galaxies which derive most of their radio emission from supermassive black holes (SMBH)\footnote{Besides a small number of galactic sources (namely PN and HII regions) the only known non-AGN in the AT20G survey is NGC253; a nearby galaxy which hosts a strong nuclear starburst \citep{ngc253}.}. On the other hand, low-frequency selected samples contain a mixture of AGN and star-forming galaxies \citep{mauch07}. Furthermore, for AGN that are observed at lower frequencies, the dominant source of emission is the large-scale radio lobes which have been built up over much longer timescales, whereas observing at high radio frequencies provides insight into the most recent activity in the central SMBH. 

Understanding the high-frequency radio population is vital to understanding the growth and evolution of AGNs and provides complementary data to their low-frequency counterparts. An analysis of the radio properties of AT20G show that it is a very different radio population from the low-frequency selected radio population. \citet{at20ganalysis} find that the catalogue is dominated by flat-spectrum sources, particularly at bright fluxes where 81\% of the sample has flat radio spectra (defined as $\alpha^{20}_8 > -0.5$ where $S_{\nu}\propto\nu^{\alpha}$) which drops of to 60\% for fluxes below 100\,mJy. A recent review by \citet{dezotti10} provides a comprehensive view of the radio properties of AGN, including both low and high-frequency information.
 
Whilst the radio data provides a unique sample of objects, these data alone are insufficient to constrain models of radio source properties and the evolution of radio galaxies. Additional optical data, particularly spectroscopic information, is vital in understanding the physical properties of the central black hole. 

There have been a number of other surveys carried out at high radio frequencies. These include; the Wilkinson Microwave Anisotropy Probe (WMAP), which covers the whole sky at 22\,GHz down to a 1\,Jy flux limit, the Planck Early Release Compact Source Catalogue \citep{planckcat, planckanalysis}, an all-sky survey at 30\,GHz, with a flux limit of 0.5\,Jy and the 9th and 10th Cambridge surveys (9C; \citealt{9c}, and 10C; \citealt{10c}) carried out at 15\,GHz that cover a much smaller area, but observe down to fainter flux limits (5.5\,mJy and 1.0\,mJy respectively). However, with the exception of \citet{bolton04} who followed up sources in the 9C sample at both optical and radio frequencies, there has not been much focus on the optical properties of these surveys.

It has been known for sometime that radio galaxies can show a range of optical spectral signatures. Some galaxies display strong emission lines whilst others have very weak or no emission features (\citealt{hine+longair} and more recently \citealt{sadler02}). In recent years these differences have been attributed to different accretion modes \citep{best05, hardcastle07}. `Cold-mode' accretion occurs when cold gas is accreted onto the central SMBH, which gives rise to the traditional AGN picture of an accretion disk surrounded by a dusty torus, a broad-line region close to the accretion disk and narrow-line region further out \citep{antonucci93}. Such objects show high-excitation lines in the optical spectrum, either narrow or broad depending on the alignment of the AGN to our line of sight. `Hot-mode' accretion occurs when hot-gas is being accreted, resulting in a radiatively inefficient accretion disk that observationally shows none of the conventional signs of AGN activity in the optical regime. It is theorised that cold-mode (also referred to as `quasar-mode') accretion occurs as a result of mergers which supply a reservoir of cold gas, whilst hot-mode (also known as `radio-mode') accretion is the result of hot gas from the surrounding ISM falling into the central core \citep{croton06}.

It was thought that these different accretion modes could account for the observed dichotomy of radio-loud AGN \citep{fanaroff}. However, it has been shown that in general the radio properties of AGN observed at 1.4\,GHz are completely independent of the optical emission line properties \citep{best05}, but are well correlated with the stellar mass of the galaxy (see also \citealt{mauch07}). By studying the optical properties of a high-frequency radio sample we aim to investigate the link between the 20\,GHz emission (predominantly from the core of the AGN) and the optical properties since both are generated on similar size and time scales. 

Additional optical and redshift information of high-frequency radio sources also provides valuable information for both the Fermi and ESA's Planck missions, which rely on multiwavelength data at higher angular resolution to correctly identify their sources. It has already been shown that there is large overlap between the AT20G survey and both the Fermi Large Area Telescope First Source Catalog ({\it Fermi} 1FGL; \citealt{fermi1FGL}) and the Early Release Compact Source Catalog \citep{planckcat}; see \citet{fermiat20g} and \citet{paco} respectively. The postional uncertainty of these instruments require multiwavelength information to accurately identify the correct counterpart. The positional accuracy of AT20G ($\sim 1$\,arcsec) allows us to identify the correct optical source and redshift, providing valuable information for these surveys.

The primary goal of this paper is to present accompanying optical and redshift data for the AT20G survey, but we also present some of the basic population statistics of a high radio-frequency selected survey. Section~\ref{makingids} details how the optical identifications were made and Section~\ref{redshifts} discusses how redshifts were found. Section~\ref{specclass} describes the spectral properties as defined from the optical spectrum and Section~\ref{catalogue} describes the resulting AT20G--optical catalogue. In Section~\ref{results} we present results from the radio--optical analysis. This is followed by a comparison of the AT20G catalogue with the S-cubed semi-empirical models \citep{s3-sex}, which make predictions of the 18\,GHz sky, in Section~\ref{skads}. We conclude in Section~\ref{conclusions}. Throughout this paper we use the following cosmological parameters: $H_0 = 71$ km s$^{-1}$ Mpc$^{-1}$, $\Omega_m=0.27$ and $\Omega_{\Lambda}=0.73$ \citep{wmap7}.

\section{Identifying the optical counterparts of AT20G sources} \label{makingids}

\begin{figure}
\epsfig{file=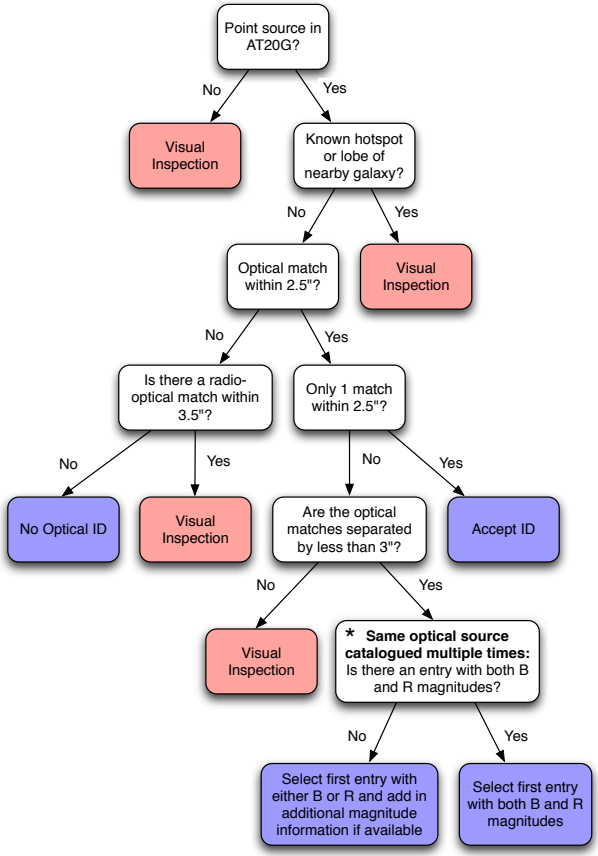, width=\linewidth}
\caption{Flowchart of automated optical identification process. The red boxes show the stages where visual inspection was neccessary and the blue boxes are when an ID was accepted. The step marked with an * is explained in more detail in Section~\ref{makingids}. \label{flowchart}}
\end{figure}

Optical identifications of AT20G sources were made using the SuperCOSMOS Science Archive \footnote{http://surveys.roe.ac.uk/ssa/} \citep{supercosmos}. AT20G sources in the Galactic plane ($|b|\leq10^{\circ}$) were excluded due to the contamination of foreground stars and dust extinction. This leaves a sample of 4932 AT20G sources for which optical IDs were sought. A semi-automated process was employed to identify the optical counterparts in an attempt to achieve a complete and reliable catalogue, while limiting the number of sources identified manually. An overview of this process is shown in Figure~\ref{flowchart}.

\begin{figure}
\epsfig{file=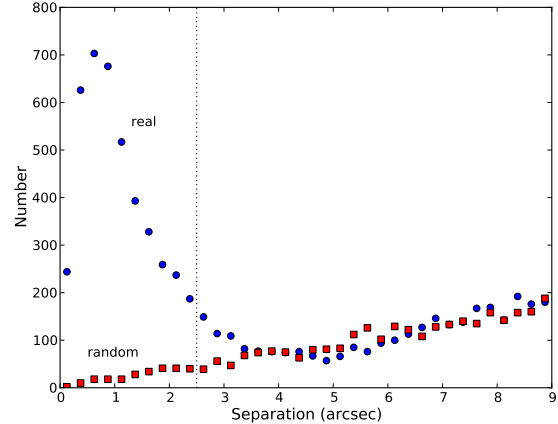, width=\linewidth}
\caption{Monte carlo test to determine the best cutoff to automatically accept optical identifications. The dashed line at 2.5\,arcsec  separation between the optical and radio positions marks the cutoff chosen. \label{montecarlo}}
\end{figure}

Due to the large number of radio point sources in the AT20G catalogue (94\%) the majority of identifications were made based on the separation between the radio and optical positions. A Monte Carlo test was carried out to determine the optimum separation at which an optical ID could be automatically accepted. A demonstration of the simulation is shown in Figure~\ref{montecarlo}. A random catalogue of radio sources was generated by offsetting the AT20G positions in declination, and optical matches found for this fake catalogue. At a separation of 3.5\,arcsec the number of real matches is equal to the number of fake matches so matches beyond this separation are unlikely to be genuine. The turnover in the number of real matches seen at very small separations is due to the position uncertainties of AT20G sources. The mean positional errors are 0.9\,arcsec in RA and 1.0\,arcsec in Dec \citep{at20g}. 

To limit the number of false matches, we automatically accepted all radio-optical matches within 2.5\,arcsec. At separations less than 2.5\,arcsec only 4.3\% of the fake AT20G catalogue had an optical match, giving a reliability of more than 95\%. For completeness, AT20G sources that had an optical counterpart between 2.5 and 3.5\,arcsec were checked manually. In many cases the optical counterpart was the correct ID and the larger offset was caused by position uncertainties either due to a weak AT20G source (where the radio positions are slightly less accurate) or due to a very bright nearby galaxy (where the optical position is more uncertain). 

\begin{table*}
\begin{center}
\caption{Examples of optical IDs that are catalogued twice in the SuperCOSMOS Science Archive, yet clearly correspond to the same object. In the first case (AT20GJ000249$-$211419) the second entry was accepted as the ID since it has the most complete record. In the second case (AT20GJ223043$-$632926), since there is no entry with both B and R magnitudes, the optical position accepted is the one with the smallest offset from the AT20G position and the first B magnitude measurement (B $=22.68$) and first R band magnitude (R $=21.05$) taken to be the correct magnitudes. The information listed here is the exact output from the SuperCOSMOS Science Archive which lists the positions in decimal degrees. \label{multmatches}}
\begin{tabular}{lccccc}
\hline
{\bf AT20Gname} &  \multicolumn{2}{c}{\bf optical position} & {\bf offset} & {\bf B} & {\bf R} \\
& RA & Dec & (arcmin) & & \\
\hline
AT20GJ000249$-$211419 & $+$0.7073508 & $-$21.2389971 & $+$0.0267763 & $-$99.999990 & $-$99.999990 \\
AT20GJ000249$-$211419 & $+$0.7072780 & $-$21.2389685 & $+$0.0289446 & $+$19.678000 & $+$19.173000 \\
AT20GJ223043$-$632926 & $+$337.6811523 & $-$63.4908370 & $+$0.0046435 & $-$99.999990 & $+$21.046000 \\
AT20GJ223043$-$632926 & $+$337.6810423 & $-$63.4909069 & $+$0.0065117 & $+$22.682000 & $-$99.999990 \\ 
\hline
\end{tabular}
\end{center}
\end{table*}
    
A number of sources had multiple optical matches within the 2.5\,arcsec search radius. Inspection of these sources revealed that they are generally the same source, but have been catalogued in the SuperCOSMOS database more than once with slightly different positions. In many of these cases one or more of the catalogued entries had no magnitude information. Examples are given in Table~\ref{multmatches}. To first confirm that the multiple entries correspond to the same source, the separation between the optical matches was calculated. If the separation was less than 3\,arcsec (corresponding to approximately 4 pixels in the SuperCOSMOS images\footnote{the sampling of the UKST plates is 10$\umu$m or 0.7\,arcsec and the typical seeing is approximately 2\,arcsec.}) it was assumed that the multiple matches refer to the same source. In this case, the most complete record (i.e. the entry with both B and R magnitude information) was accepted as the ID. If there was no catalogued source with both magnitudes listed, then the first entry with either magnitude was selected and the other magnitude added in if available in a different entry. If the separation between the multiple optical matches was larger than 3\,arcsec, visual inspection was required to confirm the correct optical ID. 

For sources that were flagged as extended in the AT20G catalogue, or sources known to be hotspots of nearby galaxies, the optical identifications were made by visual inspection. This involved creating overlays with the AT20G 20\,GHz contours and 1\,GHz contours (from either the SUMSS or NVSS surveys) overlaid on the SuperCOSMOS B-band image. The low-frequency information was needed in order to identify the optical ID where large radio lobes were present. Of the sources flagged as extended in the AT20G catalogue, approximately half were resolved into 2 separated components as determined from the visibilities. For these double sources an additional flag `d' was added in the final catalogue. 

\begin{figure*}
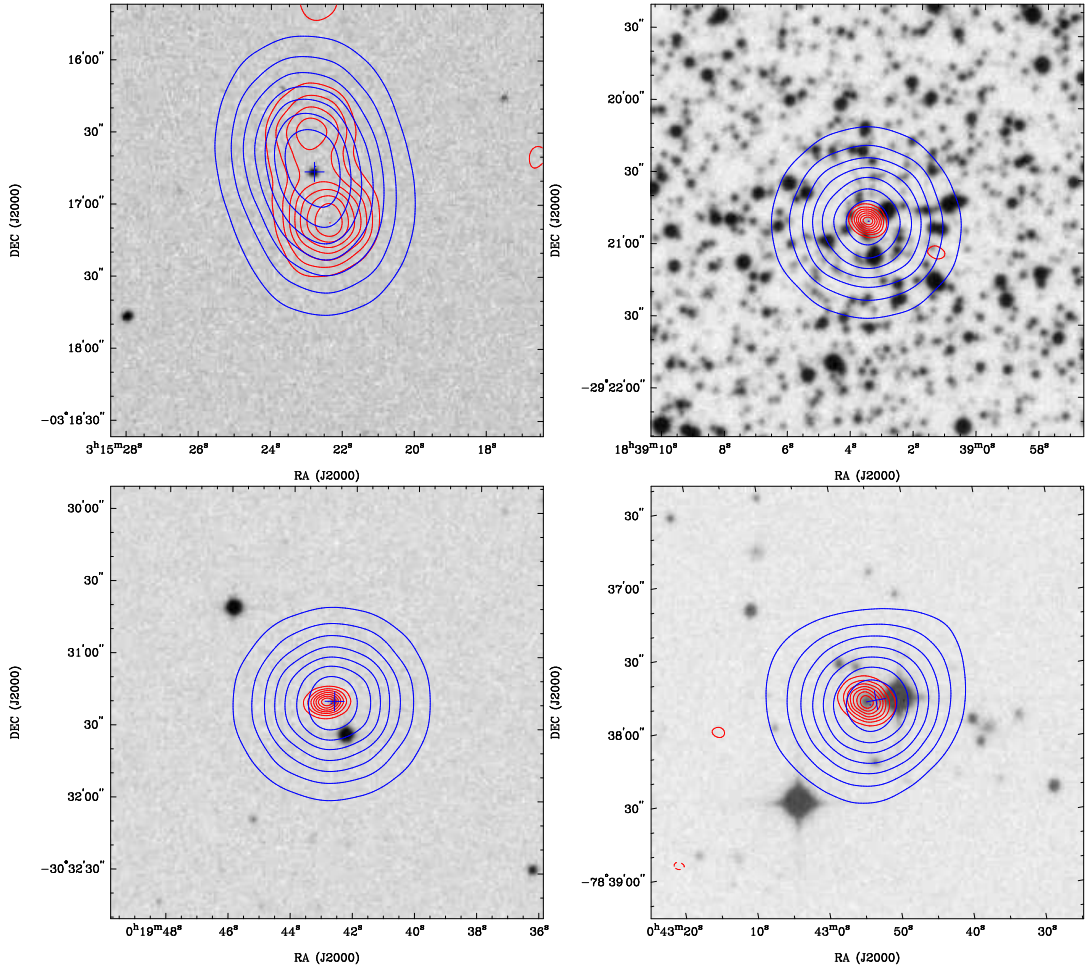

\begin{minipage}{0.4\linewidth} 
\epsfig{file=0315-0317.ps, width=0.9\linewidth, angle=270}
\end{minipage}
\begin{minipage}{0.4\linewidth}
\epsfig{file=1839-2921.ps, width=0.9\linewidth, angle=270}
\end{minipage}
\begin{minipage}{0.4\linewidth}
\epsfig{file=0019-3030.ps, width=0.9\linewidth, angle=270}
\end{minipage}
\begin{minipage}{0.4\linewidth}
\epsfig{file=0042-7838.ps, width=0.9\linewidth, angle=270}
\end{minipage}
\caption{Examples illustrating the additional flags used in the AT20G--optical catalogue. Working clockwise from the top left are examples of an object flagged as double (`d'), crowded (`c'), blended (`b') and offset (`o'). The greyscale is the SuperCOSMOS B-band image with AT20G 20\,GHz (red/inner contours) and either the NVSS or SUMSS 1\,GHz (blue/outer contours) overlaid. The optical position is denoted by the blue cross. \label{opticalflags}}
\end{figure*}

For a small subset of sources, an optical ID could not be easily determined from the overlay image. In some cases this was due to the optical field being too crowded or due to a foreground star along the line of sight of the AT20G source. This was typically a problem towards Galactic features (such as the Orion nebula) as well as in the Magellanic clouds. These crowded sources are flagged with a `c'. In other cases the optical and radio positions were slightly offset from each other, making it difficult to tell if the match is genuine or just a chance alignment. In many cases it is possible that this is due to an error with the AT20G position, so the optical information was included in the catalogue flagged with a `o'. For a small fraction of sources there were two optical sources close on the sky that were blended into the one catalogued source, yet visual inspection shows them to be two distinct sources. In many cases this was due to a foreground star so both the optical positions and magnitudes should be treated as unreliable. These blended sources have been flagged with a `b' and are excluded from the analysis where magnitudes were used. Examples of each of these types of flagged sources are shown in Figure~\ref{opticalflags}. For some of the very nearby galaxies the SuperCOSMOS magnitudes are unreliable, due to the large angular sizes. To exclude these unreliable magnitudes from the analysis, a flag `u' was added in the catalogue if the B$-$R colours were either less than $-2$ or greater than 4. One additional target (AT20GJ133608$-$082952) was also found to have unreliable magnitudes, yet was not captured by the colour cutoff. This source was also flagged with a `u' and excluded from the analysis.

\begin{figure}
\epsfig{file=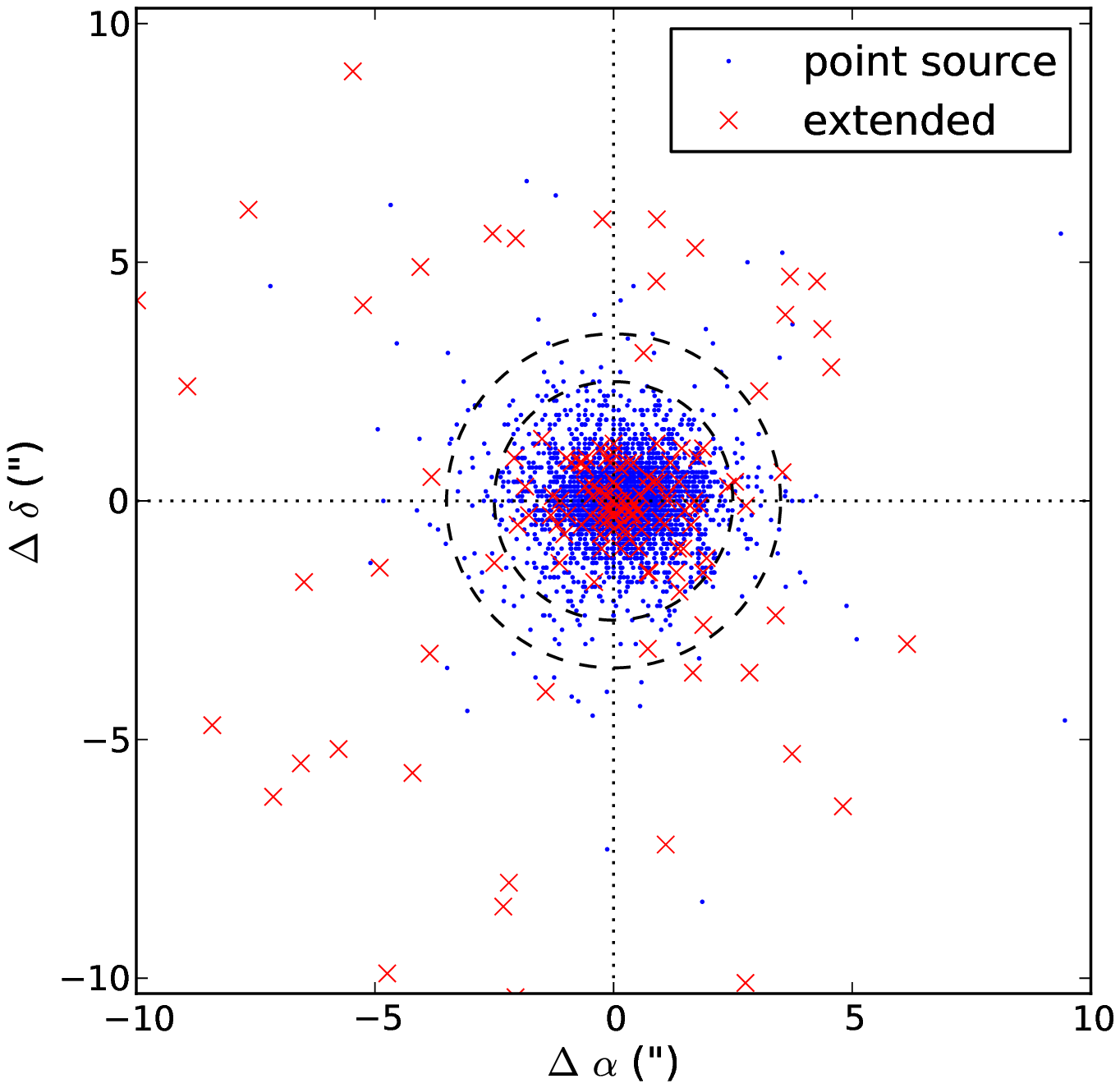, width=\linewidth}
\caption{Positional offset between the radio and optical positions for the accepted AT20G identifications. The dashed lines correspond to 2.5 and 3.5\,arcsec radii. \label{posoffset}}
\end{figure}

A total of 3873 (78.5\%) AT20G sources have optical identifications. This is much higher than what is traditionally seen at lower frequencies where the optical identification rate (to a similar magnitude limit of B=22) for radio sources selected at $\sim$1\,GHz is approximately 25--30\% \citep{bock99}. Previous AT20G papers have reported a lower optical identification rate of approximately 60\% for the AT20G survey \citep{at20g, at20ganalysis}. This is due to a simplified identification process which only accepted the sources that had matches within 2.5\,arcsec and B$\leq$22 (the SuperCOSMOS completeness limit), highlighting the incompleteness of only using automatic crossmatching procedures. This AT20G--optical catalogue provides a more complete and reliable sample of optical counterparts which was not available at the time of publication of the earlier papers. Figure~\ref{posoffset} shows the offset in RA and Dec between the optical and 20\,GHz radio position for all of the accepted optical identifications. The dashed circle represents the 2.5 arcsec radius within which sources were automatically accepted. The vast majority of optical IDs fall within this circle.

\begin{figure}
\epsfig{file=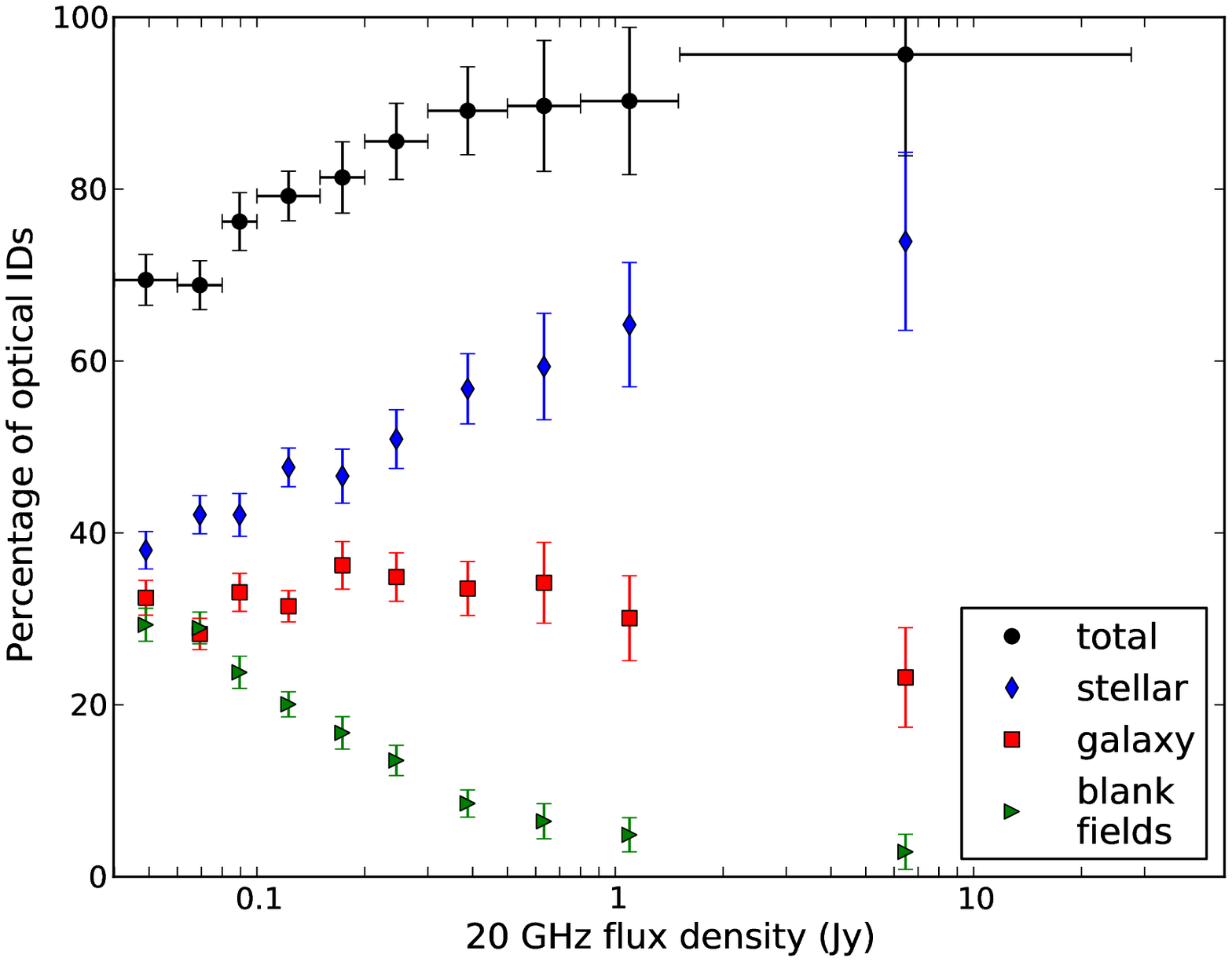, width=\linewidth}
\caption{Optical identification rate as a function of 20\,GHz flux. The black circles are for the total fraction of identified optical couterparts in each flux bin, blue diamonds are the objects classified as stellar in SuperCOSMOS (most likely to be QSOs) and the red squares are the SuperCOSMOS galaxies. The green triangles are the fraction of AT20G sources that do not have an observed optical counterpart (blank fields). The horizontal error bars denote the bins used and the vertical errorbars are the poisson errors. \label{opticalidvsflux}}
\end{figure} 

Figure~\ref{opticalidvsflux} shows that the optical identification rate increases as a function of 20\,GHz flux density. Using the SuperCOSMOS classification scheme based on the optical morphology as being point-like (termed `stellar') or resolved (`galaxy') we can see that this increase in optical identification rate is due to the increase in the fraction of stellar objects (i.e. QSOs) shown by the blue diamonds in Figure~\ref{opticalidvsflux}. On the other hand the fraction of optical IDs classified as galaxies, along with those sources that did not have ID (blank fields), which we assume are distant galaxies below the sensitivity limit of the optical plates, increases with {\it decreasing} flux density. Extrapolating from this plot, deeper high frequency surveys would have a lower identification rate as galaxies become the dominant population.

\section{Finding Redshifts} \label{redshifts}

After identifying the optical counterparts of AT20G sources, we searched for corresponding redshift information. Knowing the optical position allowed for a much smaller search radius hence limiting the number of spurious redshifts.

\subsection{Searching the 6dF Galaxy Survey} \label{6df}

The 6-degree field Galaxy Survey (6dFGS; \citealt{6df}) is a spectroscopic survey of the entire southern sky, providing an ideal sample to crossmatch with the AT20G survey. The primary 6dFGS sample is a 2MASS K-band selected sample, but there are also many additional target samples that were observed as part of the survey. For more details on the selection criteria for the additional targets see \citet{6df2004}. We searched the 6dFGS database for sources within 6\,arcsec of the optical counterpart\footnote{The diameter of the 6dF fibres is 6.7\,arcsec.}. Sources that had a radio--optical separation larger than 2.5\,arcsec were checked manually to ensure that no redshifts were missed by having a small search radius around the optical position. This was generally found to be a problem for very nearby galaxies where the optical positions were not accurate enough to warrant a 6\,arcsec search radius.

A total of 433 AT20G sources were observed as part of the 6dF Galaxy Survey. Approximately 35\% of these were within the primary K-band selected sample, while the majority of the remaining sources were part of the X-ray selected (see \citealt{rass6df}) or low-frequency radio selected \citep{mauch07} additional targets. Each spectrum obtained in 6dF was assigned a quality from 1--6; spectra of quality 1 were of poor quality and a redshift was not able to be measured. Spectra of quality 2 means that the spectrum obtained was of good quality (i.e. sufficient signal-to-noise), but a redshift was not able to be measured. Quality 2 spectra sources are typically BL-Lac objects. Redshifts of quality 3 or 4 denote a reliable redshift while quality 6 was assigned for a spectrum of a $z=0$ star. Of the sources observed with 6dF, 346 had reliable redshifts (quality 3,4 or 6). 

\subsection{Searching the Literature}

The majority of redshifts were obtained using the NASA Extragalactic Database (NED). Once again, we searched for redshifts within 6\,arcsec of the optical position, and manually checked those positions with large radio--optical offsets. A total of 1264 AT20G sources have redshifts listed in NED. For these sources a classification also accompanied the redshift (generally `G' or `QSO'). However, since many sources have no associated optical spectra it is difficult to tell how accurate these classifications are. 

A number of AT20G sources that were observed in the 6dF Galaxy Survey also have redshifts listed in NED. When compiling the AT20G--optical catalogue we placed priority on the 6dFGS redshifts since we have access to the optical data to confirm the redshift. For sources that weren't observed with 6dF or didn't have a reliable redshift, the literature redshift was used. The final catalogue contains 395 6dF sources, of which 343 have reliable redshift measurements\footnote{The three sources with good quality redshifts that did not end up in the final catalogue are dominated by a foreground star in the 6dF spectrum. The redshift of the background AT20G source is listed in NED and so this redshift is listed in the catalogue.} and 977 sources with literature redshifts. Where a 6dF redshift is used, the 6dF sourcename (of the form gHHMMSS$-$DDMMSS \citealt{6df2004}) is listed as the redshift reference. If the redshift was obtained from NED, then the corresponding bibliographic code that is listed alongside the redshift is used\footnote{An explanation of these bibcodes can be found at http://adsdoc.harvard.edu/abs\_doc/help\_pages/data.html.}

\subsection{Additional Redshifts}

To increase the redshift coverage of the sample, additional observations were carried out (see Table~ \ref{additionalobs}). Many of the AT20G sources have optically bright counterparts making them ideal back-up targets for observing in poor weather conditions. These observations resulted in 144 extra redshifts. We have also assigned qualities to each spectrum in the same way as was done for the 6dFGS sources. 

\begin{table}
\begin{center}
\caption{Summary of redshifts obtained from either the literature, 6dFGS or additional observations. The first column is the facility used for the observations followed by the number of sources observed (where applicable) and the number of sources where a redshift was able to be measured. The final column shows the redshift reference code used in the AT20G--optical catalogue.  \label{additionalobs}}
\begin{tabular}{lcccc}
\hline
{\bf Telescope} & {\bf Date of} & \multicolumn{2}{c}{\bf Targets} & {\bf Reference} \\
& {\bf obs.} & {\bf obs.} & {\bf with z} &  \\
\hline
{\bf NED:} &&& &\\
&  & --- & 977 & Bibcode \\
{\bf 6dF:} &&& & \\
UKST & 2001--2006 & 395 & 343 & 6dF:6dfname \\
{\bf other:} &&& \\
SSO 2.3m  & 2007 & 19 & 16 & SSO\_07  \\ 
SSO 2.3m  & 2008 & 42 & 39 & SSO\_08  \\ 
ESO--NTT & 2008 & 10 & 10 & NTT \\
Gemini--South & 2009--2010 & 51 & 50 & Gemini \\
ESO--NTT & 2010 & 24 & 22 & ICRF \\
SSO 2.3m & 2010 & 7 & 7 & SSO\_10 \\
\hline
\end{tabular}
\end{center}
\end{table}

\subsubsection{SSO 2.3\,m data}

There were four separate observing runs using the ANU 2.3\,m telescope located at the Siding Spring Observatory in which AT20G sources were used as back-up targets in case of poor weather. The first of these was carried out over the nights of 2007 April 11--14 by Richard Hunstead and EKM. This was followed by a second run from 2007 June 22--23, carried out by HMJ and EKM. Both of these runs have the redshift reference flag `SSO\_07'. Targets were observed using the 158R grating in the blue arm of the Dual-Beam Spectrograph (DBS) and had exposure times ranging from 10--30\,mins, depending on the magnitude of the target. The third set of targets were observed from 2008 May 1--5 by PJH and EKM using the same setup. The main targets of this particular run were AT20G selected Gigahertz Peaked Spectrum (GPS) sources described in \citet{hancock}. Additional optically bright AT20G sources were also included as back-up targets. 

During 2009 the DBS was decommissioned and replaced with the Wide-Field Spectrograph (WiFeS)\footnote{See http://www.mso.anu.edu.au/wifes/}. The final set of 2.3m redshifts were observed over 2010 September 08--10 by KWB and PJH using WiFeS with the B3000 and R3000 gratings and the RT560 dichroic. AT20G sources were again poor-weather targets and were observed for 5\,mins each. All of the 2.3\,m data was reduced using the Interactive Data Reduction and Analysis Facility (\texttt{IRAF}; \citealt{iraf}), and redshifts were measured using the \texttt{RUNZ} program (originally written by W.~Sutherland; \citealt{runz}).

\subsubsection{ESO--NTT data}

Data was obtained from the 3.6\,m European Southern Observatory (ESO) New Technology Telescope (NTT) during 2008 August 24--29 as back-up targets for a program observed by Richard Hunstead and HMJ. Targets were observed using the ESO Faint Object Spectrograph and Camera v.2 (EFOSC2) and had exposure times ranging from 400--1000\,s. Again the data was reduced using \texttt{IRAF} and \texttt{RUNZ}.

An observing program targeting southern radio sources in the International Celestial Reference Frame was carried out in August 2010 \citep{ICRF}. Since this is a radio VLBI selected sample, there was significant overlap with the AT20G catalogue. Redshifts that were obtained as part of this observing program have been included in the AT20G--optical catalogue with the reference `ICRF'.  

\subsubsection{Gemini--South data}

A number of AT20G sources were observed as part of the poor weather program on Gemini--South under program ID GS-2009B-Q-95. Poor weather program targets are classified as `band 4' and are observed when there is nothing higher in the queue that can be targetted. Sources were observed as longslit observations using the Gemini Multi-Object Spectrograph and reduced using the {\it Gemini} package in \texttt{IRAF}. As this observing program was solely focused on obtaining redshifts for AT20G sources, spectra obtained here are shown in Appendix \ref{appendixa}. 

\subsection{Redshift Completeness}

\begin{figure}
\epsfig{file=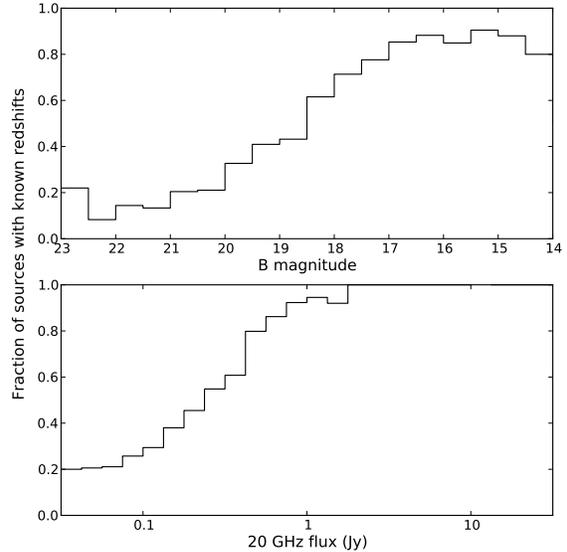, width=\linewidth}
\caption{Fraction of sources with redshifts as a function of optical magnitude (top) and 20\,GHz flux density (bottom). While we are resonably complete for bright AT20G sources, this completeness drops off significantly for fainter objects. \label{redshiftfrac}}
\end{figure}

Of the 3873 sources with optical counterparts, 1525 (39.4\%) have redshift information. This corresponds to 30.9\% of the entire AT20G sample (with $|b|>10^{\circ}$). Figure~\ref{redshiftfrac} shows the fraction of sources with redshifts as a function of both 20\,GHz flux and B magnitude. As expected, for the brighter sources the fraction of sources with redshifts is quite high ($\sim$80\%), but this drops off very rapidly with 20\,GHz flux. Ignoring known Galactic planetary nebula and HII regions, the redshift sample is 91.5\% complete for sources above 1\,Jy and 85.3\% complete above 500\,mJy.

In order to calculate the luminosity function of high-frequency sources it is vital to have a complete redshift sample, or at the very least be able to quantify any bias and selection effects. Since our redshift sample is compiled from a range of different surveys and observing runs, each with their own selection criteria, qualification of these biases and selection effects is beyond the scope of this paper. We continue to collect redshift information where available and this will be the subject of future papers. The local ($z<0.15$) 20\,GHz luminosity function using redshifts from the 6dF Galaxy Survey is the subject of a forthcoming paper (Sadler et al. 2011, in preparation).

\section{Spectral Classifications} \label{specclass}

For AT20G objects that had an associated optical spectrum (from either 6dFGS or follow-up observations) we classified the spectrum according to their emission or absorption features. This subset of objects form the `spectroscopic sample'. Sources that had reliable redshift measurements (those with qualities greater than 2) were classified into the following categories, based on \citet{mauch07};\\

\noindent{\bf Aa:} AGN with only absorption features\\
{\bf Aae:} AGN with both absorption and emission lines present\\
{\bf Ae:} AGN with only emission features\\
{\bf AeB:} AGN with broad emission lines\\
{\bf SF:} Star-forming galaxy\\
{\bf Star:} Spectrum of a star \\

\begin{figure*}
\epsfig{file=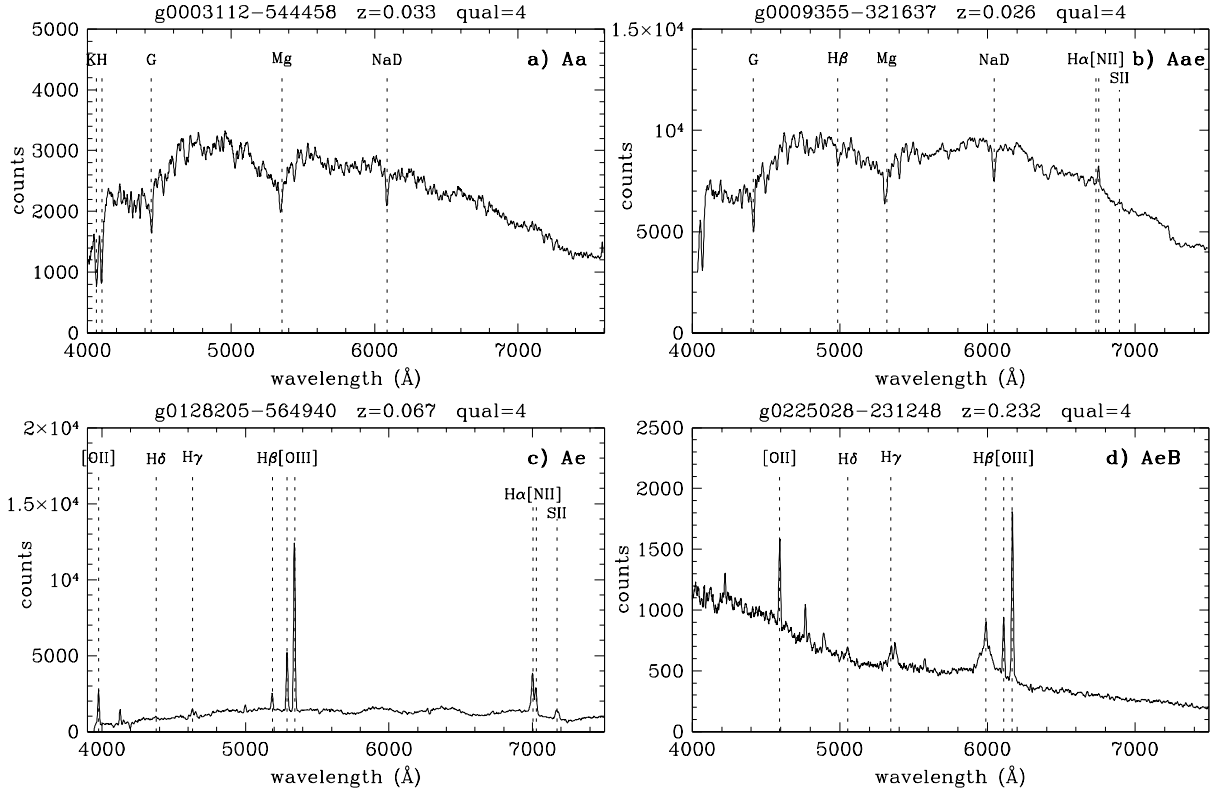, width=\linewidth}
\caption{Examples of the different spectral classifications. From the top left the different classes are Aa (absorption line only), Aae (absorption and weak emission lines), Ae (strong emission lines) and AeB (broad emission lines). The spectra shown here have been smoothed using a boxcar algorithm over $10 {\rm{\AA}}$. \label{examplespectra}}
\end{figure*}

\noindent An example spectrum for each of the Aa, Aae, Ae and AeB spectral classes is shown in Figure~\ref{examplespectra}. In addition to these, a {\bf BL-Lac} classification was given to sources exhibiting a featureless spectrum with high continuum, particularly in the blue, which were generally assigned a quality of `2'.

Table~\ref{samples} lists the different subsets of AT20G sources and the number of sources that fall into each category. To investigate how well the spectroscopic sample represents the full AT20G population, Table~\ref{samples} also lists the median 20\,GHz flux densities, spectral indices, B magnitudes, redshifts, radio luminosity and absolute B magnitudes. As already seen in Figure~\ref{redshiftfrac} the redshift sample (all objects that have redshift information, including those from NED) is dominated by the brighter sources at both radio and optical wavelengths. As such, the median 20\,GHz flux, radio luminosity and B magnitude are all slightly higher than that of the full AT20G sample, or those that have optical IDs. The spectroscopic sample forms a subset of the redshift sample, hence are also dominated by the brighter AT20G sources. However, Table~\ref{samples} shows that the median 20\,GHz flux density is lower than that of the redshift sample, and more like the median flux seen for the full AT20G survey. This is most likely due to the fact that many of the brighter radio sources have literature redshifts and therefore weren't followed-up spectroscopically in either the 6dF survey or other additional observations. 

The spectroscopic sample is biased towards the optically brighter sources and, due primarily to the large number of spectra obtained from the 6dF survey, is dominated by lower redshift targets. Where the spectral properties are used in the following analysis it should be noted that results are  representative of the optically brighter AT20G objects and may not neccessarily apply to the fainter targets. 

\begin{table*}
\begin{center}
\caption{The different subsamples of the AT20G--optical catalogue discussed throughout this paper. The number of sources comprising each sample is shown in the 2nd row, followed by the median values of a range of radio and optical properties. The parentheses denote the standard errors of the median. \label{samples}}
\begin{tabular}{lcccc}
\hline
&{\bf AT20G} & {\bf Optical} & {\bf Redshift} & {\bf Spectroscopic} \\ 
\multicolumn{2}{r}{($|b|>10^{\circ}$)} & {\bf IDs} & {\bf Sample} & {\bf Sample} \\
& & & (incl. NED) & (no NED $z$'s)\\
\hline
{\bf No.} & 4932 & 3873 & 1525 & 548 \\
{\bf $\langle$S$_{20} \rangle$ } (mJy) & 104.0 {\scriptsize($11.6$)} & 112.0 {\scriptsize($14.6$)} & 179.0 {\scriptsize($36.0$)} & 117.5 {\scriptsize($42.4$)} \\
{\bf $\langle \alpha^{20}_5\rangle$} & $-$0.22 {\scriptsize($0.01$)} & $-$0.19 {\scriptsize($0.01$)} & $-$0.23 {\scriptsize($0.02$)} & $-$0.17 {\scriptsize($0.03$)}\\
{\bf $\langle$B mag$\rangle$} & --- & 19.57 {\scriptsize($0.04$)} & 18.53  {\scriptsize($0.07$)}& 17.75 {\scriptsize($0.10$)}\\
{\bf $\langle$z$\rangle$} & --- & --- & 0.798 {\scriptsize($0.03$)} & 0.447 {\scriptsize($0.04$)} \\
{\bf $\langle$log P$_{20} \rangle$} (W/Hz) & --- & --- & 26.59 {\scriptsize($0.04$)} & 25.84 {\scriptsize($0.08$)} \\
{\bf $\langle$M$_{\rm B} \rangle$} & --- & --- & $-$24.43 {\scriptsize($0.36$)} & $-$23.72 {\scriptsize($0.14$)} \\
\hline
\end{tabular}
\end{center}
\end{table*}

\subsection{The AT20G spectroscopic sample}

\begin{figure}
\epsfig{file=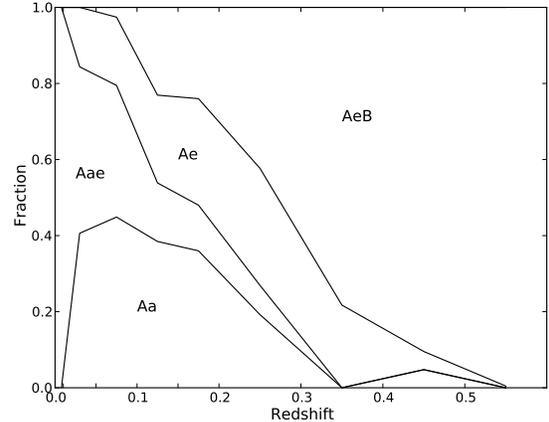, width=\linewidth}
\caption{The fraction of each spectral class as a function of redshift. The data are binned in redshift and the solid lines show the contribution of each spectral classification to the total population. \label{specvsz}}
\end{figure}

Figure~\ref{specvsz} shows the fraction of each class observed as a function of redshift. The `Aae' class is only observed at low redshifts due to the high signal-noise that is required to be able to distinguish weak emission lines. In addition, aperture effects mean that with increasing redshift more light from the surrounding galaxy is observed in the fixed-aperture fibre. This extra galaxy light dominates the spectrum, potentially overpowering any weak emission lines from the nuclear region. At redshifts larger than $z=0.5$ the sample is comprised completely of `AeB' objects. All other classes fall out of the sample by this redshift either because the spectral features are too weak to distinguish (such as the `Aa' and `Aae' classes) or the optical magnitude is so faint that they are not detected as optical ids (as is the case for the `Ae' class). There are no star-forming galaxies detected in the spectroscopic AT20G sample, yet there is at least one known starburst (NGC253) in the full AT20G catalogue. 

\begin{figure}
\epsfig{file=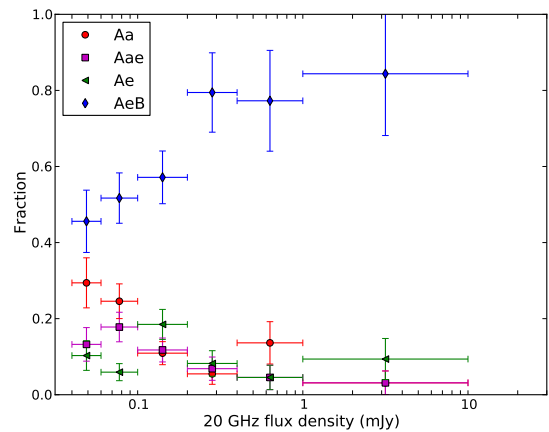, width=\linewidth}
\caption{The fraction of each spectral class binned in 20\,GHz flux density. The horizontal error bars denote the bin widths and the vertical error bars are the poisson errors. \label{specvsflux}}
\end{figure}

Figure~\ref{specvsflux} shows the fraction of each spectral class as a function of 20\,GHz flux. The majority of sources fall into either the `Aa' or `AeB' categories so it is difficult to investigate any trends for the `Aae' and `Ae' classes. The fraction of `AeB' type sources increases with flux, similar to the increase of `stellar' optical sources seen in Figure~\ref{opticalidvsflux}. Similarly the fraction of `Aa' sources decreases with flux in the same way that `galaxy' optical IDs do in Figure~\ref{opticalidvsflux}.

\subsection{SuperCOSMOS classifications vs. spectral classifications}

\begin{figure}
\epsfig{file=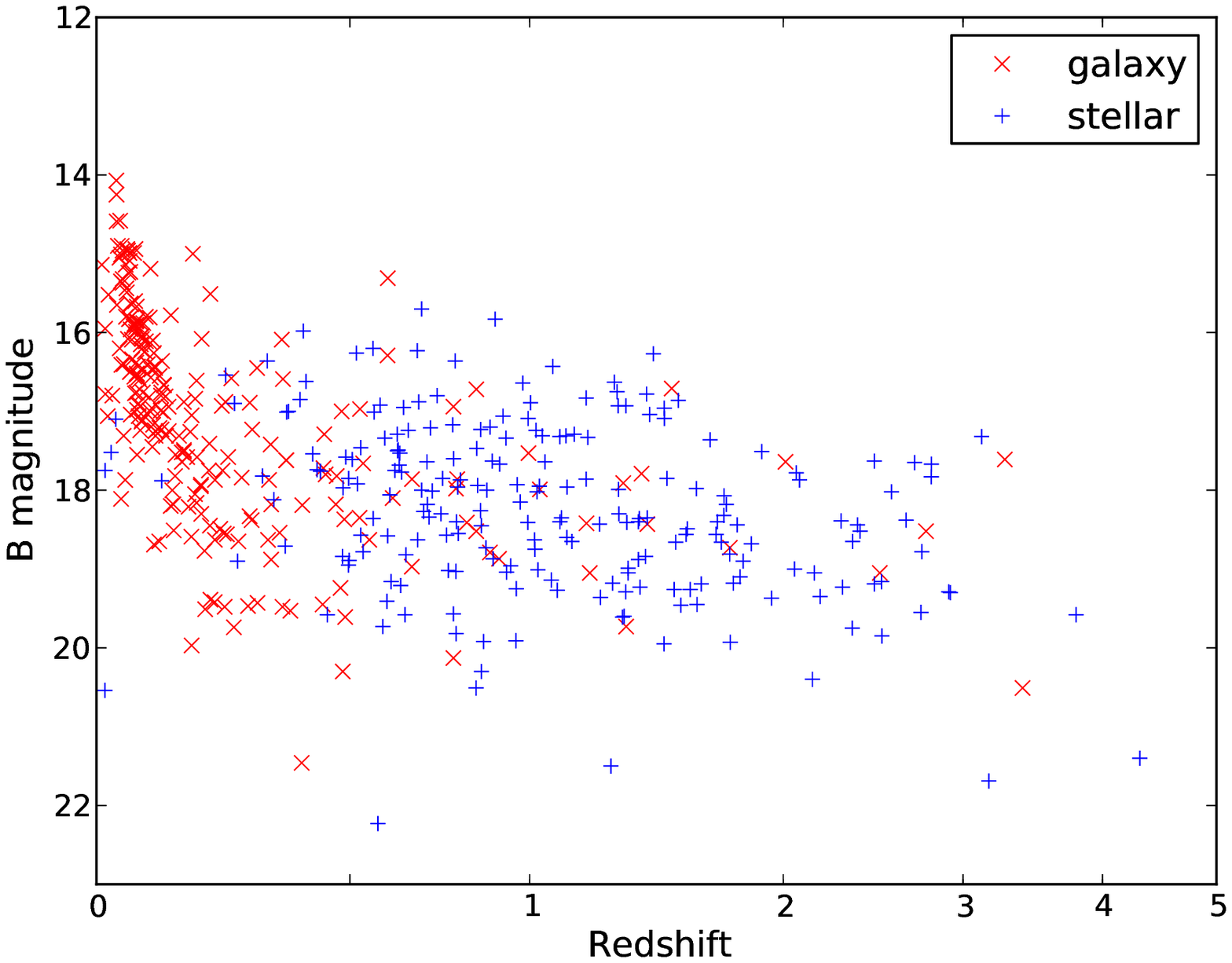, width=\linewidth}
\caption{B magnitude against redshift (plotted as log($1+z$)) for all AT20G sources with a redshift. The sample is separated according to the SuperCOSMOS classifications (i.e. stellar or galaxy) based on the optical morphology of the source. \label{bmagvsz}}
\end{figure}

The subset of sources with spectral classifications allowed us to check the reliability of the SuperCOSMOS classifications (`stellar' or `galaxy') which are based on the optical morphology. Figure~\ref{bmagvsz} shows the relationship between B magnitude and redshift for all AT20G sources in the redshift sample, separated into the two SuperCOSMOS classes. While most of the galaxies follow the trend of decreasing magnitude with redshift, falling below the sensitivity limit (and therefore out of the sample) at a redshift of $z\sim$ 0.5, a number of sources do not. These appear to be more aligned with the stellar sources indicating that the optical IDs were misclassified by SuperCOSMOS. 

\begin{figure}
\epsfig{file=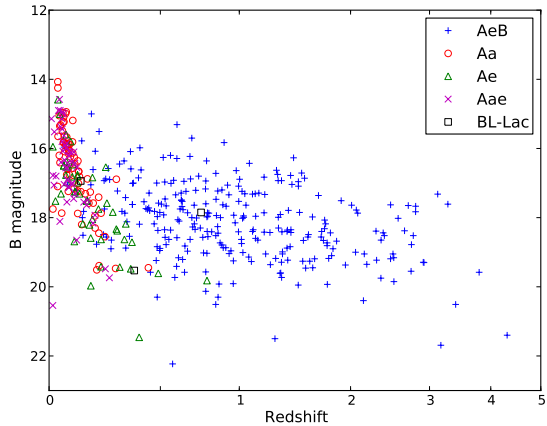, width=\linewidth}
\caption{B magnitude against redshift (plotted as log($1+z$)) for sources in the spectroscopic sample. The different spectral classifications are plotted to compare with the SuperCOSMOS classifications shown in Figure~\ref{bmagvsz}. \label{zbmagspec}}
\end{figure}

This misclassification can be confirmed by comparing Figure~\ref{bmagvsz} with Figure~\ref{zbmagspec}, which shows the relationship between B magnitude and redshift for the different spectral classifications. Objects classified as `Aa', `Aae' or `Ae' fall into the SuperCOSMOS `galaxy' category and `AeB' sources are generally associated with `stellar' optical IDs as expected. However, 27\% of sources that are classed as galaxies by SuperCOSMOS exhibit broad emission lines (AeB spectral type). While it is possible that some of these sources could be nearby Type 1 Seyfert galaxies, most of the AeB sources (70\%) are above $z=0.3$ further indicating that they are probably QSOs.

\begin{figure}
\epsfig{file=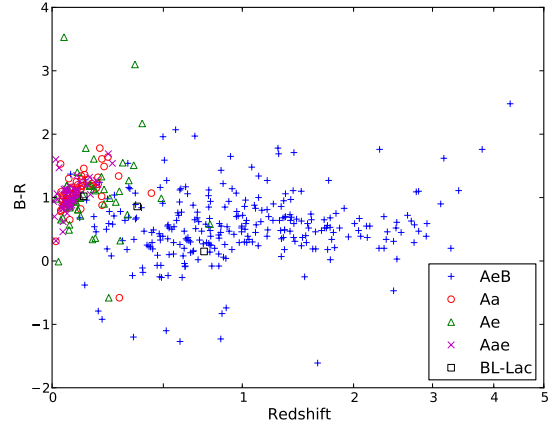, width=\linewidth}
\caption{B$-$R colours against redshift(plotted as log($1+z$)) for sources in the spectroscopic sample that have reliable redshift measurements. \label{zcolourspec}}
\end{figure}

The relationship between optical colours (B$-$R) and redshift is shown in Figure~\ref{zcolourspec}. Again the sample is separated into two populations; the galaxies become redder with increasing redshift until they fall out of the sample at $z\sim0.5$, whereas QSOs are generally bluer, but remain flat across a wide range of redshifts. The upturn seen at $z>3$ is due to the Lyman-$\alpha$ forest moving into the B-band at these redshifts.

\section{The AT20G--Optical Catalogue} \label{catalogue}

The AT20G--optical catalogue is available as supplementary information in the online version of this paper. The catalogue consists of the following columns:\\

\noindent(1) AT20Gname: In the format JHHMMSS$-$DDMMSS corresponding to the source position in J2000 coordinates.  \\
(2--3) RA and Dec of the AT20G position.\\
(4--6) AT20G 20, 8.6 and 4.8\,GHz flux measurements from the AT20G catalogue in units of mJy. \\
(7) 1.4\,GHz flux from the NRAO VLA Sky Survey (NVSS) in mJy. \\
(8) 843\,MHz flux from the Sydney University Molonglo Sky Survey (SUMSS) in mJy. \\ 
(9) Quality flag as listed in the AT20G catalogue. This is either `g' or `p' and relates to the quality of the flux measurement as being good or poor \citep{at20g}. \\
(10) Other AT20G flags as listed in the AT20G catalogue. These are presented in \citet{at20g}, but are also listed below. An additional flag `d' has been added to denote sources that have multiple components in the radio as discussed in Section~\ref{makingids}. \\
(11--12) RA and Dec of the optical counterpart. The positions were taken from the SuperCOSMOS Science Archive. \\   
(13) Offset between the AT20G and optical position in arcsec. \\
(14--15) $b_{\rm J}$ and $r_{\rm F}$ magnitudes from the SuperCOSMOS Science Archive. \\
(16) SuperCOSMOS class based on the morphology of the optical source. A value of 1 means the source was classified as a `galaxy', while a value of 2 denotes a `stellar' (i.e. point-like) source. Values greater than 2 indicate that the source was unable to be classified.\\
(17) Optical flags as described in Section~\ref{makingids}.\\
(18--19) Redshift and quality measurement. Qualities of 3 or 4 denote a reliable redshift measurement, while values less then 3 means that the source was observed, but no redshift was able to be measured. For these sources the redshift is given as $-$9.99. A quality `N' means the redshift was obtained through NED. \\
(20) Spectral classification. See Section~\ref{specclass} for the different classifications used. For redshifts obtained from the literature the NED classification (either `G', `GPair' or `QSO') is listed. \\  
(21) Absolute magnitude calculated from the SuperCOSMOS $r_{\rm F}$ magnitudes and shifted into the rest frame (see Section~\ref{kcorr}). \\
(22) Logarithm of the K-corrected 20\,GHz Radio luminosity. In units of W/Hz. \\
(23) Reference for redshift as listed in Table~\ref{additionalobs}.\\

The flags listed in Column 10 denote the following:\\
{\bf e:} AT20G source is extended. \\
{\bf d:} AT20G source has multiple components (usually a double). \\
{\bf h:} Galactic HII region. \\
{\bf p:} Planetary Nebula. \\
{\bf m:} In the Magellanic Clouds. \\
{\bf l:} No low-frequency counterpart (see \citealt{at20g} for further details). 

\begin{landscape}
\pagestyle{empty} 
\topmargin=+1.5cm 
\oddsidemargin=-2.0cm
\evensidemargin=-1.5cm

\begin{table} 
\scriptsize{
 \begin{center}
    \begin{minipage}{\linewidth}
\caption{The first page of the AT20G--optical catalogue. The full catalogue is available in the electronic version of the journal.
 \label{examplecat}}
   \newcolumntype{V}{>{\centering\arraybackslash} m{0.7cm} }
   \newcolumntype{W}{>{\centering\arraybackslash} m{0.1cm} }   
   \newcolumntype{X}{>{\centering\arraybackslash} m{0.3cm} }   
    \begin{tabular}{{lccXXXWWccccXccWWcWXccc}}
\hline
{\bf AT20Gname} & {\bf RA} & {\bf Dec} & {\bf S20} & {\bf S8.4} & {\bf S4.8} & {\bf Q} & {\bf E} & {\bf S1.4} & {\bf S0.8} & \multicolumn{2}{c}{\bf Optical Position} (J2000) & {\bf offset} & {\bf B} & {\bf R} & {\bf SC} & {\bf flag} & {\bf z} & {\bf q} & {\bf spec.} & {\bf M$_{\rm R}$} & {\bf log P$_{20}$}  & {\bf reference} \\
& \multicolumn{2}{c}{(J2000)} & (mJy) & (mJy)& (mJy)& & &(mJy) &(mJy) & {\bf RA} & {\bf Dec} & (") & mag & mag & {\bf cl.} & & & & {\bf class} & & (W/Hz) &  \\ 
\hline
AT20GJ000012$-$853919 & 00:00:12.78 & $-$85:39:19.9 & 98 & 63 & 63 & g & . & - & 104.5 & 00:00:13.64 & $-$85:39:21.4 & 1.9 & 19.89 & $-$99.9 & 1 & b & - & - & - & - & - & - \\ 
AT20GJ000020$-$322101 & 00:00:20.38 & $-$32:21:01.2 & 118 & 315 & 515 & g & . & 520.9 & 321.5 & 00:00:20.38 & $-$32:21:01.2 & 0.1 & 18.70 & 18.34 & 2 & . & 1.275 & N & QSO & $-$25.98 & 27.07 & 1989QSO...M...0000H \\ 
AT20GJ000105$-$155107 & 00:01:05.42 & $-$15:51:07.2 & 297 & 295 & 257 & g & . & 347.5 & - & 00:01:05.33 & $-$15:51:07.0 & 1.3 & 18.16 & 18.39 & 1 & . & 2.044 & N & QSO & $-$27.05 & 27.44 & 1990PASP..102.1235T \\ 
AT20GJ000106$-$174126 & 00:01:06.31 & $-$17:41:26.2 & 73 & - & - & g & . & 447.4 & - & 00:01:06.28 & $-$17:41:26.7 & 0.7 & 22.64 & $-$99.9 & 2 & . & - & - & - & - & - & - \\ 
AT20GJ000118$-$074626 & 00:01:18.04 & $-$07:46:26.8 & 177 & - & - & g & . & 208.4 & - & 00:01:18.02 & $-$07:46:26.9 & 0.3 & 18.05 & 17.09 & 2 & . & $-$9.99 & 1 & - & - & - & 6dF:g0001180$-$074627 \\ 
AT20GJ000124$-$043759 & 00:01:24.50 & $-$04:37:59.6 & 50 & - & - & g & . & 632.3 & - & 00:01:24.64 & $-$04:38:00.0 & 2.1 & 22.16 & 20.73 & 2 & . & - & - & - & - & - & - \\ 
AT20GJ000125$-$065624 & 00:01:25.59 & $-$06:56:24.7 & 77 & - & - & g & . & 58.0 & - & 00:01:25.58 & $-$06:56:24.9 & 0.3 & 19.31 & 18.63 & 2 & . & - & - & - & - & - & - \\ 
AT20GJ000212$-$215309 & 00:02:12.02 & $-$21:53:09.9 & 165 & - & - & g & . & 370.5 & - & 00:02:11.97 & $-$21:53:09.7 & 0.7 & 19.14 & 18.55 & 1 & . & - & - & - & - & - & - \\ 
AT20GJ000221$-$140643 & 00:02:21.71 & $-$14:06:43.9 & 48 & - & - & g & . & 823.5 & - & - & - & - & - & - & - & . & - & - & - & - & - & - \\ 
AT20GJ000230$-$033140 & 00:02:30.60 & $-$03:31:40.1 & 53 & - & - & g & . & 67.9 & - & 00:02:30.62 & $-$03:31:40.4 & 0.4 & 18.40 & 17.68 & 2 & . & - & - & - & - & - & - \\ 
AT20GJ000249$-$211419 & 00:02:49.85 & $-$21:14:19.2 & 100 & - & - & g & . & 122.2 & - & 00:02:49.75 & $-$21:14:20.3 & 1.7 & 19.68 & 19.17 & 1 & . & - & - & - & - & - & - \\ 
AT20GJ000252$-$594814 & 00:02:52.93 & $-$59:48:14.0 & 71 & 64 & 57 & g & . & - & 62.1 & 00:02:52.96 & $-$59:48:14.6 & 0.6 & $-$99.9 & 21.59 & 2 & . & - & - & - & - & - & - \\ 
AT20GJ000253$-$562110 & 00:02:53.65 & $-$56:21:10.8 & 94 & 229 & 403 & g & . & - & 1350.0 & 00:02:53.57 & $-$56:21:10.9 & 0.7 & 22.99 & 20.44 & 2 & . & - & - & - & - & - & - \\ 
AT20GJ000303$-$553007 & 00:03:03.45 & $-$55:30:07.1 & 44 & 48 & 52 & g & . & - & 44.2 & 00:03:03.33 & $-$55:30:07.2 & 1.0 & 22.20 & 20.66 & 2 & . & - & - & - & - & - & - \\ 
AT20GJ000311$-$544516 & 00:03:11.04 & $-$54:45:16.8 & 95 & 313 & 552 & g & ed & - & 1549.0 & 00:03:10.55 & $-$54:44:56.2 & 21.1 & 14.25 & 13.45 & 1 & . & 0.033 & 4 & Aa & $-$22.33 & 23.36 & 6dF:g0003112$-$544458 \\ 
AT20GJ000313$-$590547 & 00:03:13.33 & $-$59:05:47.7 & 49 & 101 & 151 & g & . & - & 505.9 & 00:03:13.33 & $-$59:05:47.7 & 0.1 & 19.21 & 19.51 & 2 & . & - & - & - & - & - & - \\ 
AT20GJ000316$-$194150 & 00:03:16.06 & $-$19:41:50.7 & 76 & 162 & 182 & g & . & 232.3 & - & 00:03:15.94 & $-$19:41:50.3 & 1.7 & 19.26 & 18.75 & 2 & . & - & - & - & - & - & - \\ 
AT20GJ000322$-$172711 & 00:03:22.05 & $-$17:27:11.9 & 386 & - & - & g & . & 2414.8 & - & 00:03:22.11 & $-$17:27:14.1 & 2.4 & 16.93 & 16.17 & 1 & . & 1.465 & N & QSO & $-$28.48 & 27.33 & 1978MNRAS.185..149H \\ 
AT20GJ000327$-$154705 & 00:03:27.35 & $-$15:47:05.4 & 129 & - & - & g & . & 527.3 & - & 00:03:27.26 & $-$15:47:05.7 & 1.4 & 22.57 & 21.01 & 2 & . & 0.508 & N & G & $-$21.61 & 25.93 & 2002A\&A...386...97J \\ 
AT20GJ000404$-$114858 & 00:04:04.88 & $-$11:48:58.0 & 680 & - & - & g & . & 459.2 & - & 00:04:04.90 & $-$11:48:58.4 & 0.5 & 18.07 & 18.62 & 2 & . & -9.99 & 2 & bllac & - & - & 6dF:g0004049$-$114858 \\ 
AT20GJ000407$-$434510 & 00:04:07.24 & $-$43:45:10.0 & 199 & 211 & 244 & g & . & - & 343.8 & 00:04:07.24 & $-$43:45:10.0 & 0.1 & 18.53 & 18.28 & 2 & . & - & - & - & - & - & - \\ 
AT20GJ000413$-$525458 & 00:04:13.97 & $-$52:54:58.7 & 65 & 98 & 192 & g & e & - & 276.9 & 00:04:14.00 & $-$52:54:58.5 & 0.3 & 14.07 & 13.29 & 1 & . & 0.032 & 4 & Aa & $-$22.47 & 23.18 & SSO\_10 \\ 
AT20GJ000435$-$473619 & 00:04:35.65 & $-$47:36:19.0 & 868 & 970 & 900 & g & . & - & 932.9 & 00:04:35.64 & $-$47:36:19.5 & 0.4 & 17.63 & 17.60 & 2 & . & 0.884 & 4 & AeB & $-$25.83 & 27.26 & Gemini \\ 
AT20GJ000505$-$344549 & 00:05:05.94 & $-$34:45:49.6 & 131 & 142 & 134 & g & . & 91.7 & 80.9 & 00:05:05.95 & $-$34:45:49.6 & 0.1 & 21.83 & -99.9 & 2 & . & - & - & - & - & - & - \\ 
AT20GJ000507$-$013244 & 00:05:07.03 & $-$01:32:44.6 & 81 & - & - & g & . & 70.7 & - & 00:05:07.06 & $-$01:32:45.2 & 0.8 & 18.66 & 18.31 & 1 & . & 1.710 & N & QSO & $-$26.71 & 26.78 & 1995AJ....109.1498H \\ 
AT20GJ000518$-$164804 & 00:05:18.01 & $-$16:48:04.9 & 142 & - & - & g & . & 261.8 & - & 00:05:17.93 & $-$16:48:04.7 & 1.2 & 17.98 & 17.73 & 1 & . & 0.777 & 3 & AeB & $-$25.39 & 26.35 & 6dF:g0005179$-$164805 \\ 
AT20GJ000558$-$562828 & 00:05:58.32 & $-$56:28:28.9 & 151 & 376 & 677 & g & ed & - & 2763.0 & 00:05:57.91 & $-$56:28:31.3 & 4.1 & 19.27 & 17.46 & 1 & . & 0.291 & N & G & $-$23.79 & 25.61 & 20032dF...C...0000C \\ 
AT20GJ000600$-$313215 & 00:06:00.47 & $-$31:32:15.0 & 63 & 53 & 52 & g & . & 44.3 & 48.3 & 00:06:00.37 & $-$31:32:14.8 & 1.3 & 21.10 & 19.90 & 1 & . & - & - & - & - & - & - \\ 
AT20GJ000601$-$295549 & 00:06:01.14 & $-$29:55:49.6 & 97 & 187 & 228 & g & . & 87.5 & - & 00:06:01.12 & $-$29:55:50.0 & 0.6 & 18.15 & 18.82 & 2 & . & - & - & - & - & - & - \\ 
AT20GJ000601$-$423439 & 00:06:01.95 & $-$42:34:39.8 & 110 & 259 & 532 & g & . & - & 2850.0 & 00:06:01.86 & $-$42:34:40.8 & 1.5 & 20.65 & 19.74 & 1 & . & 0.530 & N & G & $-$22.70 & 26.10 & 2006AJ....131..114B \\ 
AT20GJ000610$-$830543 & 00:06:10.18 & $-$83:05:43.5 & 167 & 361 & 691 & g & e & - & 2548.0 & - & - & - & - & - & - & . & - & - & - & - & - & - \\ 
AT20GJ000613$-$062334 & 00:06:13.90 & $-$06:23:34.8 & 2084 & - & - & g & . & 2050.7 & - & 00:06:13.89 & $-$06:23:35.3 & 0.5 & 19.48 & 17.98 & 1 & . & 0.347 & 4 & Ae & $-$23.63 & 26.79 & 6dF:g0006139$-$062335 \\ 
AT20GJ000619$-$424518 & 00:06:19.72 & $-$42:45:18.3 & 183 & 189 & 199 & g & . & - & 246.0 & 00:06:19.72 & $-$42:45:18.5 & 0.1 & 18.27 & 18.40 & 2 & . & 1.770 & N & QSO & $-$26.70 & 27.19 & 1977ApJ...217..362H \\ 
AT20GJ000622$-$000423 & 00:06:22.59 & $-$00:04:23.2 & 425 & - & - & g & . & 3897.6 & - & 00:06:22.60 & $-$00:04:24.5 & 1.3 & 20.02 & 19.71 & 2 & . & 1.037 & N & QSO & $-$24.11 & 27.08 & 1985PASP...97..932S \\ 
AT20GJ000635$-$731144 & 00:06:35.21 & $-$73:11:44.7 & 51 & 54 & 54 & g & . & - & 63.4 & - & - & - & - & - & - & . & - & - & - & - & - & - \\ 
AT20GJ000713$-$402337 & 00:07:13.41 & $-$40:23:37.4 & 69 & 89 & 97 & g & . & - & 98.5 & 00:07:13.42 & $-$40:23:37.4 & 0.1 & $-$99.9 & 19.78 & 2 & . & - & - & - & - & - & - \\ 
AT20GJ000720$-$611306 & 00:07:20.56 & $-$61:13:06.7 & 150 & 138 & 134 & g & . & - & 85.8 & 00:07:20.55 & $-$61:13:06.7 & 0.1 & 19.92 & 20.75 & 2 & . & 0.857 & 4 & AeB & $-$22.61 & 26.44 & Gemini \\ 
AT20GJ000800$-$233918 & 00:08:00.42 & $-$23:39:18.0 & 154 & - & - & g & . & 374.5 & - & 00:08:00.36 & $-$23:39:18.3 & 0.9 & 16.78 & 16.41 & 2 & . & 1.411 & 4 & AeB & $-$28.16 & 26.90 & 6dF:g0008004$-$233918 \\ 
AT20GJ000801$-$524339 & 00:08:01.71 & $-$52:43:39.9 & 124 & 127 & 101 & g & . & - & 255.8 & 00:08:01.67 & $-$52:43:39.8 & 0.5 & 18.91 & 19.45 & 1 & . & - & - & - & - & - & - \\ 
AT20GJ000809$-$394522 & 00:08:09.18 & $-$39:45:22.9 & 146 & 169 & 188 & g & . & 150.1 & 134.2 & 00:08:09.20 & $-$39:45:23.0 & 0.2 & 19.76 & 19.89 & 2 & . & - & - & - & - & - & - \\ 
AT20GJ000826$-$255911 & 00:08:26.27 & $-$25:59:11.2 & 120 & 259 & 360 & g & . & 468.2 & - & 00:08:26.10 & $-$25:59:11.3 & 2.3 & $-$99.9 & 19.96 & 2 & . & - & - & - & - & - & - \\ 
AT20GJ000828$-$132930 & 00:08:28.02 & $-$13:29:30.3 & 161 & - & - & g & . & 191.7 & - & - & - & - & - & - & - & . & - & - & - & - & - & - \\ 
AT20GJ000829$-$055845 & 00:08:29.33 & $-$05:58:45.2 & 66 & - & - & g & . & 1321.9 & - & 00:08:29.28 & $-$05:58:45.7 & 0.9 & 19.58 & 18.11 & 1 & . & - & - & - & - & - & - \\ 
AT20GJ000831$-$141959 & 00:08:31.08 & $-$14:19:59.9 & 53 & - & - & g & . & 603.5 & - & - & - & - & - & - & - & . & - & - & - & - & - & - \\ 
AT20GJ000837$-$461940 & 00:08:37.53 & $-$46:19:40.8 & 75 & 98 & 95 & g & . & - & 113.8 & 00:08:37.54 & $-$46:19:40.7 & 0.2 & 17.42 & 16.57 & 2 & . & 1.850 & N & QSO & $-$28.64 & 26.89 & 1993RMxAA..25...51M \\ 
AT20GJ000922$-$513011 & 00:09:22.06 & $-$51:30:11.9 & 49 & 59 & 56 & g & . & - & 45.1 & 00:09:22.03 & $-$51:30:11.6 & 0.4 & 16.81 & 15.62 & 1 & . & 0.117 & 4 & Aa & $-$23.17 & 24.19 & 6dF:g0009220$-$513012 \\ 
AT20GJ000927$-$161028 & 00:09:27.55 & $-$16:10:28.8 & 69 & 92 & 61 & g & . & 55.0 & - & 00:09:27.50 & $-$16:10:29.2 & 0.8 & 20.86 & 20.57 & 4 & . & - & - & - & - & - & - \\ 
AT20GJ000935$-$321637 & 00:09:35.67 & $-$32:16:37.3 & 221 & 275 & 313 & g & . & 388.3 & 558.8 & 00:09:35.71 & $-$32:16:36.1 & 1.2 & 16.80 & 15.33 & 1 & . & 0.026 & 4 & Aae & $-$19.94 & 23.50 & 6dF:g0009355$-$321637 \\ 
AT20GJ000937$-$635734 & 00:09:37.40 & $-$63:57:34.2 & 59 & 66 & 75 & g & . & - & 101.9 & 00:09:37.51 & $-$63:57:33.8 & 0.8 & 20.51 & $-$99.9 & 2 & . & - & - & - & - & - & - \\ 
\hline    
\end{tabular}
\end{minipage}
  \end{center}
}
\end{table}
\end{landscape}

\noindent {\bf b:} Source part of the AT20G `Big and Bright' sample presented in \citet{at20gb+b}.

The flags listed in Column 17 refer to the optical counterparts as discussed in Section~\ref{makingids}. In summary they indicate the following: \\
{\bf c:} Optical field is too crowded to make an identification. \\
{\bf o:} Optical source is slightly offset from the radio source, but still within 3.5\,arcsec. \\
{\bf b:} Blended source in the optical. \\
{\bf u:} Unreliable optical magnitudes. \\

The first page of the catalogue is shown in Table~\ref{examplecat}. 

\subsection{K-corrections} \label{kcorr}

To compare luminosities across a wide range of redshifts it is essential to shift the observed frequencies into the rest-frame using a k-correction. As QSOs and galaxies have very different SEDs in the optical regime we have performed different corrections for each class. For galaxies in our sample (those with spectral classifications of Aa, Aae and Ae or listed as G, GPair in NED) we applied the k-correction given in \citet{Depropris04}:

\begin{equation}
K(r_{\rm F})=((-0.08+1.45x)y)-((2.88+0.48x)y^2) %##from De Propris et al 2004
\end{equation}
where
\begin{equation}
x=b_{\rm J}-r_{\rm F}
\end{equation}

\begin{equation}
y=\frac{z}{1+z}
\end{equation}

For the sources classified as QSOs (AeB or BLLac for sources with spectral classifications or classed as QSO in NED) we assumed a power law fit for the K-correction:
\begin{equation}
K(r_{\rm F})=-2.5(1+\beta)\log(1+z)
\end{equation}
where we assume a spectral index of $\beta=-0.5$.

To calculate the rest-frame radio luminosities we applied the following k-correction for all sources:

\begin{equation}
k(z)=(1+z)^{-(1+\alpha)}
\end{equation}
where $S\propto \nu^{\alpha}$ and $\alpha$ is the spectral index from 5--20\,GHz. If there was no 5\,GHz flux available then a flat spectral index ($\alpha=0$) was assumed.

\section{Results} \label{results}

Radio sources are often divided into two populations based on their spectral indices. Flat spectrum sources (typically defined as $\alpha>-0.5$ where $S_{\nu}\propto\nu^{\alpha}$) arise from synchrotron self-absorption in the optically thick region close to the core of the AGN, and are therefore generally compact radio sources. Steep spectrum sources are dominated by the synchrotron spectrum of the optically thin lobes and hence generally correspond to extended sources. However, there are some exceptions to this rule; one such example being Compact Steep Spectrum (CSS) radio sources. 

In a similar way, optical sources are often classified into different populations using a range of diagnostics. These distinctions are often based on the morphology (i.e. QSO vs. galaxies) or on the presence or absence of emission lines. By identifying the optical counterparts of the majority of AT20G sources we aim to investigate whether the different observed radio properties can be explained by differences in the optical properties. It is particularly interesting to do this at high radio frequencies as this provides insight into the most recent activity near the core of the AGN. Sources observed at lower frequencies are dominated by emission from the radio lobes built up over long timescales and are therefore not indicative of recent accretion.

In particular, we aim to study the accretion properties of AT20G sources as defined using the optical spectra. Objects with weak or no emission lines have been classified as undergoing `hot-mode' accretion, whereas sources with strong emission lines (both broad and narrow) are in the `cold-mode' accretion phase \citep{hardcastle07}. This provides a physical interpretation of the different populations observed rather than an observational effect, and can provide insight into whether there are any intrinsic differences in the flat and steep-spectrum radio populations. 

\subsection{Optical properties of AT20G sources}

\begin{figure}
\epsfig{file=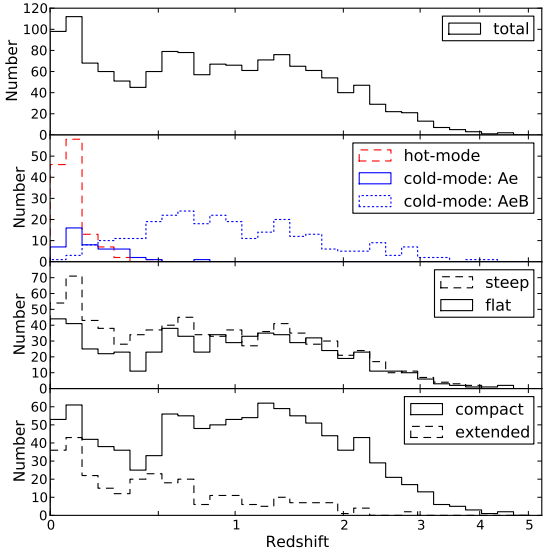, width=\linewidth}
\caption{Redshift distributions of AT20G sources binned in log($1+z$). All sources with measured redshifts are shown in the top panel, followed by sources classifed as hot or cold mode accretors, flat or steep radio spectra and compact or extended at 20\,GHz. \label{zhist}}
\end{figure}

\begin{figure}
\epsfig{file=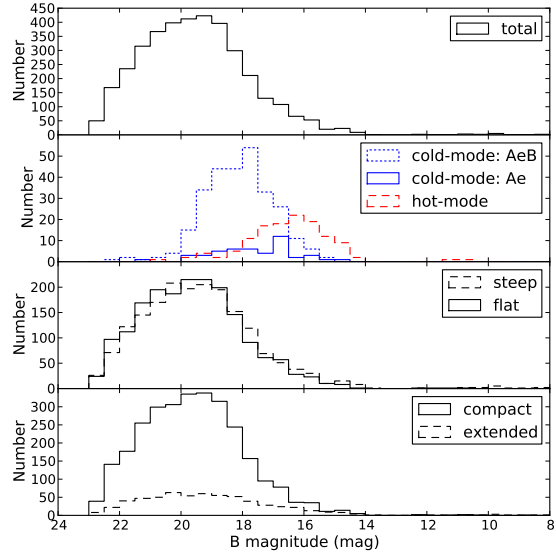, width=\linewidth}
\caption{B magnitude distribution of AT20G sources. \label{bmaghist}}
\end{figure}

Figures \ref{zhist} and \ref{bmaghist} show the redshift and B magnitude distributions of the AT20G catalogue. Each figure shows the total distribution in the top panel followed by the same distribution separated into different subpopulations based on the observed optical or radio properties. The separation into hot or cold-mode accretion was based on the emission line properties as discussed above. Sources with spectral classifications of `Aa' or `Aae' fall into the hot-mode category, while sources classified as either `Ae' or `AeB' are cold-mode. The dividing line between flat and steep radio spectra is defined as $\alpha^{20}_5>-0.5$, and the separation into compact and extended sources at 20\,GHz is based on the ratio of flux observed on the longest ($\sim$4.5~km) ATCA baselines to the short baselines. This is explained in more detail in Section~\ref{compactext}.

The redshift distribution (shown in Figure~\ref{zhist}) has a median redshift of $z=0.798$. As expected, the low redshift sources are dominated by objects with weak or no emission lines in the optical spectrum (i.e. hot-mode) while the cold-mode accretors cover a much wider range in redshift.  Separating the cold-mode accretors into the narrow and broad emission line classifications (Ae and AeB) shows that it is primarily the AeB sources that are observed out to large redshifts. Hot-mode sources (Aa and Aae) have a median redshift of $z=0.068$, while the cold-mode sources have median redshifts of $z=0.113$ for the Ae spectral class, and $z=0.877$ for the AeB class. There is a small difference observed in the redshift distributions between the flat and steep spectrum sources at lower redshifts, but beyond a redshift of $z\sim0.5$ they are very similar. A larger difference is seen between the compact and extended sources, with many more compact sources observed out to higher redshifts.

Figure~\ref{bmaghist} shows the B magnitude distribution which has a median B magnitude of B $=19.52$. Once again there is a clear difference between the hot and cold mode populations with medians of B $=16.35$ for the hot-mode sources and B $=17.31$ and B $=18.07$ for the Ae and AeB spectral classes. This is most likely a selection effect introduced by the 6dFGS sources. The majority of sources in the spectroscopic sample were observed as part of the 6dFGS, a large fraction of which were selected to be bright in the K-band. There is no obvious difference in the B magnitude distributions between the flat and steep-spectrum sources, or compact and extended sources (besides the different sample sizes). Instead of restricting the distributions to only the spectroscopic sample, we have plotted all sources where the relevant information (i.e. B magnitude or redshift) was available. As a result, the total number of sources shown in the hot vs. cold-mode distributions (which is only the spectroscopic sample) is less than the total number of sources in both the flat vs. steep and compact vs. extended distributions. 

\begin{figure}
\epsfig{file=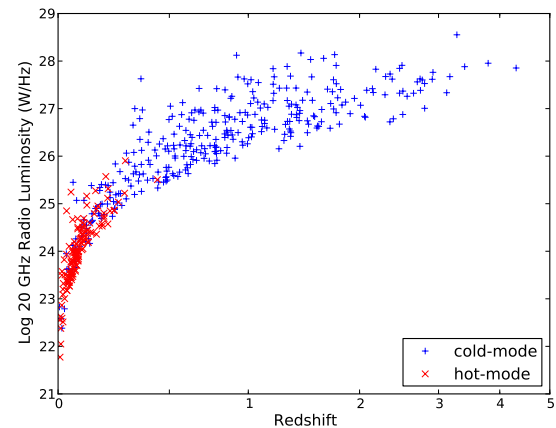, width=\linewidth}
\caption{Radio luminosity at 20\,GHz against redshift (plotted as log($1+z$)) for sources in the AT20G spectroscopic sample. Sources at higher redshifts must have higher luminosities to remain in the flux-limited sample. \label{rlumz}}
\end{figure}

\begin{figure}
\epsfig{file=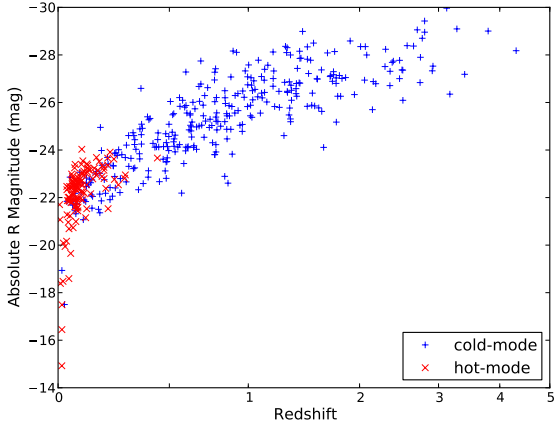, width=\linewidth}
\caption{Absolute B magnitude against redshift (plotted as log($1+z$)). As in Figure~\ref{rlumz} sources at higher redshifts have brighter absolute magnitudes as this is a flux-limited sample. \label{absmagz}}
\end{figure}

This strong selection effect on redshift can be seen in Figures \ref{rlumz} and \ref{absmagz} which plots the rest-frame 20\,GHz luminosity and absolute B magnitude against redshift. Hot-mode sources (shown by the red `x') generally have lower radio luminosities and therefore are only seen at lower redshifts. The same effect can be seen in Figure~\ref{absmagz} where objects at higher redshifts have to be intrinsically brighter to be observed above the flux limit of the survey. 

Although the radio properties of the AT20G survey are complete, there are many selection effects associated with studying the optical properties, particularly in the spectroscopic sample. To address this, when comparing the two accretion modes we have selected sources within a smaller redshift range ($z<0.3$) throughout the analysis.

\subsection{Compact vs. extended radio sources} \label{compactext}

The AT20G survey was carried out on the Australia Telescope Compact Array; a 6 element interferometer located in northern N.S.W., Australia. As described in \citet{at20g}, the observing mode used for the survey comprised 5 antennas on short baselines in a hybrid configuration and the 6th antenna on a baseline of 4.5~km. The primary data products in the AT20G catalogue were derived from the 5 dishes in a compact configuration, but the additional data from the 6th antenna is the subject of a forthcoming paper (Chhetri et al. 2011, in preparation). This additional data on the longest baselines provides information on the source structure at a resolution of $\sim$0.15\,arcsec, approximately ten times higher than the resolution obtained from the short baselines.

An approximate measure of the source size can be calculated from the ratio of flux measured on the longest baselines compared to the flux measured on the short baselines. Flux ratios greater than 0.86 correspond to the source being `compact' on the longest baselines and sources with flux ratios less than 0.86 (i.e. significantly more flux on shorter baselines) have been termed `extended' for the purpose of this analysis. The value of 0.86 was determined by looking at the scatter of values above a flux ratio of 1, which we assume is just due to gaussian noise, such that any values within that scatter ($1\pm0.14$) can be classified as compact. We have used this criteria for all of the distributions shown in Section~\ref{results}. 

\subsection{Flat vs. steep-spectrum radio sources} \label{flatvssteep}

Studying the emission line properties (hot vs. cold-mode accretion) or the radio source size (compact vs. extended) can be misleading, as both properties are strongly dependent on redshift and our spectroscopic sample is not complete. On the other hand, the radio spectral indices observed are not strongly affected by redshift and also provide information on the physical properties of the source. In particular, the distinction between flat and steep spectrum sources introduced by \citet{wall77} separates the sources dominated by a compact radio component from those dominated by more diffuse jets and lobes. Simultaneous spectral indices are available for the majority of AT20G sources south of $\delta=-15^{\circ}$, providing radio properties for a complete sample of sources which is representative of the full high-frequency radio population.

In this analysis we use the simultaneous measurement of the spectral index between 5 and 20\,GHz. Conventionally, the separation of flat and steep spectrum sources is defined by the spectral index around 1\,GHz as opposed to the relatively high frequency spectral index that we are using here. As such, compact GPS type sources may have spectra which are becoming steeper in this frequency range.

Interestingly, it is the comparison between flat and steep-spectrum populations which seem to show the least difference in Figures \ref{zhist} and \ref{bmaghist}. A Kolmogorov-Smirnov 2-sample test applied on these distributions results in a test statistic (the maximum distance between cumulative distributions) of d=0.084 for the redshift distribution and d=0.053 for the B magnitude distribution. However, due to the large number of objects in each sample, the probability that the flat and steep-spectrum sources are drawn from separate parent populations is 98.5\% for the redshift distribution and 99.2\% for the B magnitude distribution.

\begin{table}
\begin{center}
\caption{Fraction of flat vs. steep spectrum sources for each spectral class. Not all sources have 5 and 8\,GHz information, and hence were unable to be classed as either flat or steep spectrum, so the percentages have been normalised to add to 100. The brackets denote the standard error of the median. \label{flatsteepclass}}
   \newcolumntype{X}{>{\centering\arraybackslash} m{0.5cm} }   
\begin{tabular}{ccXXXXcc}
\hline
{\bf Spectral} & {\bf No.} & {\bf \%} & {\bf flat} & {\bf steep} & \multicolumn{2}{c}{\bf median}  \\
{\bf  Class.} & & & \% & \% & {\bf z} & {\bf $\alpha_5^{20}$}\\
\hline
total & 548 & 100.0 & 79.0 & 21.0 & 0.447 & $-$0.17 {\scriptsize($0.03$)}\\  
\multicolumn{3}{l}{\bf hot-mode accretors:} & & & & \\
Aa & 80 & 14.6 & 72.4 & 27.6 & 0.074 & $-$0.17 {\scriptsize($0.09$)}\\
Aae & 58 & 10.6 & 83.3 & 16.7 & 0.062 & $-$0.20 {\scriptsize($0.07$)}\\
\multicolumn{3}{l}{\bf cold-mode accretors:} & & & & \\
Ae & 50 & 9.1 & 39.3 & 60.7 & 0.113 & $-$0.69 {\scriptsize($0.11$)}\\
AeB & 280 & 51.1 & 83.4 & 16.6 & 0.877 & $-$0.15 {\scriptsize($0.04$)}\\
\multicolumn{3}{l}{\bf others:} & & & & & \\
BL-Lac & 26 & 4.8 & 100 & 0.0 & --- & $-$0.02 {\scriptsize($0.08$)}\\
Star & 16 & 2.9 & 81.8 & 18.1 & --- & --- \\
unknown & 38 & 6.9 & 84.2 & 15.8 & --- & $-$0.17 {\scriptsize($0.08$)}\\
\hline
NED z's & 977 & --- & 67.2 & 32.8 & 1.01 & $-$0.28 {\scriptsize($0.02$)}\\
optical IDs & 2348 & --- & 78.7 & 21.3 & --- & $-$0.17 {\scriptsize($0.01$)}\\ 
(without z)  & & & & & & \\
blank fields & 1059 & --- & 52.3 & 47.7 & --- & $-$0.46 {\scriptsize($0.03$)}\\
\hline
Full AT20G & 4932 & --- & 71.3 & 28.7 & --- & $-$0.22 {\scriptsize($0.01$)}\\
\multicolumn{3}{l}{(with $|b|>10^{\circ}$)} & & & & \\
\hline
\end{tabular}
\end{center}
\end{table}

To further investigate any differences between flat and steep spectrum sources, we have calculated the fraction of each for the different spectral classifications as shown in Table~\ref{flatsteepclass}. For objects in the hot accretion mode (Aa and Aae sources) the ratio of flat to steep spectrum sources is approximately the same as seen for the full spectroscopic sample (roughly 80\% flat, 20\% steep). However, for sources undergoing cold-mode accretion (Ae and AeB sources) there is a clear difference between each spectral type. Sources with broad emission lines (AeB) are predominately flat spectrum and are observed out to larger redshifts. On the other hand, narrow-line Ae objects have a larger fraction of steep spectrum sources and a lower median redshift. 

\begin{figure}
\epsfig{file=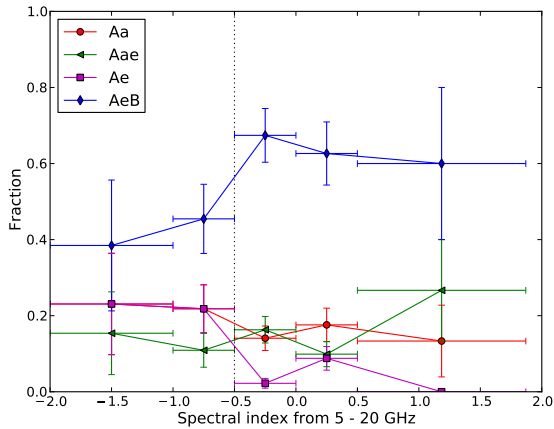, width=\linewidth}
\caption{Fraction of each spectral class observed binned in spectral index. The dashed line shows the division between flat and steep-spectrum sources at $\alpha=-0.5$ with steep sources being to the left. \label{specclassalpha}}
\end{figure} 

These differences in spectral indices can also be seen in Figure~\ref{specclassalpha} which plots the fraction of each spectral class as a function of spectral index. The hot-mode accreting sources are evenly spread across all spectral indices whereas there is a shift from flat to steep spectrum sources for the cold-mode accreting sources. AeB sources are the dominant population across all spectral indices, primarily due to the fact that these sources can be observed out to much larger redshifts, but the fraction decreases at steeper spectral indices, while the fraction of Ae sources increases at steep spectral indices.

The difference in the fraction of flat and steep-spectrum objects for the Ae and AeB spectral classes can be attributed to orientation effects in the AGN unification model \citep{antonucci93, urry+padovani}. Cold-mode accreting objects are thought to fit the canonical AGN picture with the accretion disk surrounded by a dusty torus which obscures the broad line region unless viewed pole-on. We therefore only observe AeB sources when the jets are aligned close to our line of sight, meaning that the radio emission comes predominately from the flat-spectrum, compact core. Due to this orientation effect it is likely that the core radio emission is also being Doppler boosted thereby increasing the flux. However, Ae sources are the result of larger angles of inclination, meaning that the flat spectrum core is less visible and the radio flux is dominated by the steep spectrum jet. Conversely, hot-mode accreting sources are thought to be unable to form a dusty torus or broad-line region due to inefficient accretion implying that there are no orientation differences observed between the Aa and Aae spectral classes. 

\begin{table}
\begin{center}
\caption{Fraction of flat and steep spectrum sources for sources with $z<0.3$. This allows us to compare the different spectral classifications in the same redshift range and eliminate any redshift dependancies. There is 1 additional BL-Lac object with $z<0.3$ which has not been listed in this table and accounts for 0.5\% of the total. \label{flatsteepclassz03}}
\begin{tabular}{ccccccc}
\hline
{\bf Spectral} & {\bf No.} & {\bf \%} & {\bf flat} & {\bf steep} & \multicolumn{2}{c}{\bf median}  \\
{\bf  Class.} &  & & \% & \% & {\bf z} & {\bf $\alpha_5^{20}$}\\
\hline
total & 205 & 99.5 & 70.1 & 29.9 & 0.078 & $-$0.24 {\scriptsize($0.05$)}\\
\multicolumn{3}{l}{\bf hot-mode accretors:} & & & & \\
All & 137 & 66.8 & 77.1 & 22.9 & 0.068 & $-$0.18 {\scriptsize($0.06$)}\\
Aa & 79 & 38.5 & 71.9 & 28.1 & 0.074 & $-$0.13 {\scriptsize($0.09$)}\\
Aae & 58 & 28.3 & 83.3 & 16.7 & 0.062 & $-$0.20 {\scriptsize($0.07$)}\\
\multicolumn{3}{l}{\bf cold-mode accretors:} & & & & \\
All & 67 & 32.7 & 52.4 & 47.6 & 0.134 & $-$0.49 {\scriptsize($0.11$)}\\
Ae & 43 & 21.0 & 37.5 & 62.5 & 0.099 & $-$0.77 {\scriptsize($0.12$)}\\
AeB & 24 & 11.7 & 72.2 & 27.8 & 0.194 & $-$0.10 {\scriptsize($0.14$)}\\
\hline
\end{tabular}
\end{center}
\end{table}

As noted earlier, the median redshift of Ae sources ($z=0.113$) is significantly different from the median redshift of AeB sources ($z=0.877$). To ensure that the observed differences are not simply due to redshift effects, we have also calculated the fraction of flat and steep sources for objects with $z<0.3$ as shown in Table~\ref{flatsteepclassz03}. The majority of AeB sources are removed by applying this cutoff, but the other spectral classes remain fairly similar to that in Table~\ref{flatsteepclass}. The AeB spectral class is still dominated by flat-spectrum sources and the Ae class is dominated by steep-spectrum sources. Although the number of AeB sources is much less in this redshift range, the sample is still large enough that the dominance of flat-spectrum sources is significant. We also see that the narrow emission line (Ae) sources outnumber the broad emission line (AeB) sources approximately 2:1. However, due to the incompleteness of the spectroscopic sample it is unclear whether this is indicative of the full AT20G population. 

Restricting the sample to a smaller redshift range also allows us to compare the hot and cold accretion modes with each other. As the hot-mode sources generally have lower radio luminosities, we observe a higher fraction of these sources at lower redshifts. This implies that if we were to observe to fainter flux limits these sources would become the dominant population across a much larger range in redshift. The median spectral index for the cold-mode sources ($\alpha_5^{20}=-0.49$) is steeper than the median spectral index for the hot-mode sources ($\alpha_5^{20}=-0.18$). Assuming that both populations have a core (flat-spectrum) and jet/lobe (steep-spectrum) component, the difference in spectral index implies that the cold mode accreting sources have stronger radio jets and lobes, leading to a steeper spectral index overall. This is consistent with the idea that cold-mode sources are more efficient accretors and therefore produce more powerful radio jets and lobes \citep{hardcastle07}. 

However, due to the selection effects and biases associated with the AT20G spectroscopic sample, care must be taken when interpreting these results. Separating sources into hot and cold accretion modes requires analysis of the optical spectrum and this was only available for sources observed as part of the 6dF Galaxy Survey or from additional observations. While we noted that our redshift sample was close to complete for the bright radio sources in Section~\ref{redshifts}, this is not true for the spectroscopic sample. Many of the brighter radio sources are well known objects that have literature redshifts, hence are not included in the spectroscopic sample unless they were also observed in the 6dFGS.

To address the potential biases associated with only studying the spectroscopic sample, we have also included the fraction of flat and steep spectrum sources for different subsamples of the AT20G catalogue shown at the bottom of Table~\ref{flatsteepclass}. Sources that have redshifts from NED have a slightly higher fraction of steep-spectrum sources, perhaps suggesting that a larger number of Ae type objects fall into this category, while objects that have an optical identification in SuperCOSMOS, but no redshift information, have a very similar distribution between flat and steep-spectrum sources as the spectroscopic sample. 

\begin{figure}
\epsfig{file=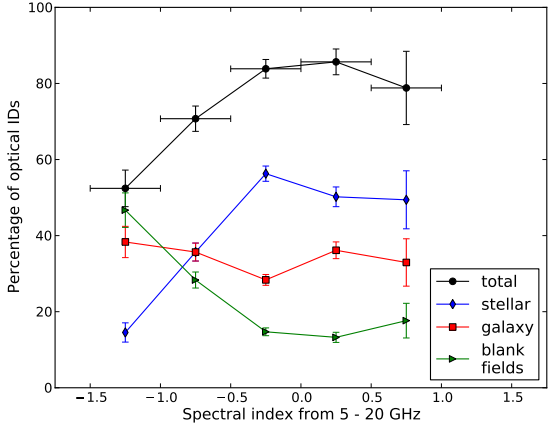, width=\linewidth}
\caption{Optical identification rate as a function of spectral index. The black circles are for the total fraction of optical IDs in each flux bin, blue diamonds are the stellar IDs (as classified in SuperCOSMOS) and the red squares are galaxies. The green triangles denote the fraction of AT20G sources that do not have an optical counterpart (blank fields). The vertical errorbars are the poisson errors and the horizontal errorbars show the bin widths. \label{opticalidvsalpha}}
\end{figure} 

\subsubsection{Unidentified sources}

Interestingly, the fraction of flat and steep spectrum sources for the blank fields is very different to that observed in the spectroscopic sample. This is also seen in Figure~\ref{opticalidvsalpha} which shows the percentage of AT20G sources that have optical IDs binned in spectral index, in a similar way to that shown in Figure~\ref{opticalidvsflux}. The galaxy IDs (red squares) show no dependence on spectral index, but the stellar optical IDs (most likely QSOs, shown by the blue diamonds) are predominately flat spectrum and drop off quite rapidly for spectral indices steeper than $-0.5$. Due to this, the total fraction of optical identifications is also lower for steep spectrum sources. This means that the percentage of blank fields increases dramatically at steep spectral indices. Many of these blank fields are likely to fall into the Ae spectral classification since this is the only category dominated by steep spectrum objects. Exhibiting only narrow emission lines means that no additional UV--optical emission from the accretion disk can be seen (unlike the AeB spectral class which can be observed to much higher redshifts) and hence the source would fall below the SuperCOSMOS plate limit by a redshift of approximately $z=0.5$. At this redshift, the radio source would have to be intrinsicially more luminous to remain in the AT20G sample and more luminous radio sources often have stronger emission lines \citep{hine+longair,rawlings+saunders,mccarthy93}. 

The increase in the percentage of blank fields at steep spectral indices could indicate a population of high-$z$ ultra-steep spectrum (USS) sources, similar to the low-frequency population first identified by \citet{blumenthal+miley}. There are 80 AT20G sources with $\alpha<-1.2$, of which 36 have no optical counterpart in the SuperCOSMOS images and potentially belong to the same class as these high redshift galaxies. These sources are listed in Appendix \ref{appendixb} along with the spectral index between 1\,GHz and 5\,GHz. Interestingly all of the sources selected to be ultra-steep between 5 and 20\,GHz have flatter spectral indices between 1 and 5\,GHz suggesting that the spectrum is curved and not a straight power-law. These sources would not have been selected as USS sources at lower frequencies, but at higher redshift they will have a steep spectrum in the GHz range.

It has been suggested that a relationship exists between the peak of the radio SED and the age of the source \citep{snellen00} such that as the source ages the spectral peak moves to lower frequencies. According to this model, USS sources selected at higher frequencies are therefore likely to be younger than those selected at lower frequencies. Appendix \ref{appendixb} lists the high-frequency USS sources with redshifts from NED where available. 

\subsection{Accretion mode vs. radio properties}

\begin{figure}
\epsfig{file=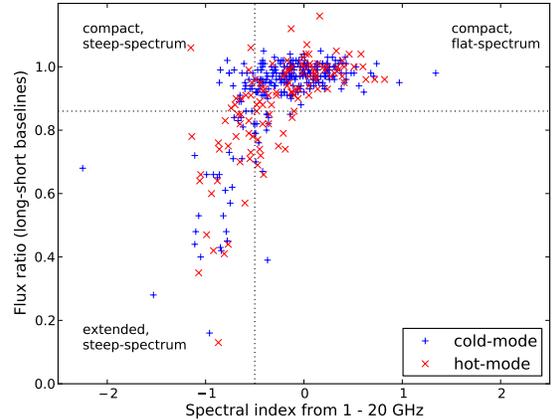, width=\linewidth}
\caption{Ratio of flux observed on the longest baselines to short baselines against spectral index for AT20G sources in the spectroscopic sample. The source structure refers to whether the source is compact or extended and is determined by the flux ratios observed on the longest ($\sim$4.5~km) ATCA baselines compared to the short baselines. The dashed lines denote the dividing lines used to separate the sample into either compact or extended source structure (flux ratio$=0.86$) and flat or steep radio spectra ($\alpha=-0.5$). \label{6kvis}}
\end{figure}

To further investigate any connection between the observed radio and optical properties we can summarise both the radio spectrum (i.e. flat vs. steep) and source size (compact vs. extended) in the one figure as shown in Figure~\ref{6kvis}. Here the flux ratio on the long--short baselines is shown against spectral index. Since many AT20G sources don't have 5\,GHz information we have used the spectral index between either 1.4 or 0.8\,GHz from the NVSS or SUMSS surveys and the AT20G 20\,GHz fluxes to obtain a larger sample. 

\begin{figure*}
\begin{minipage}{0.5\linewidth}
\epsfig{file=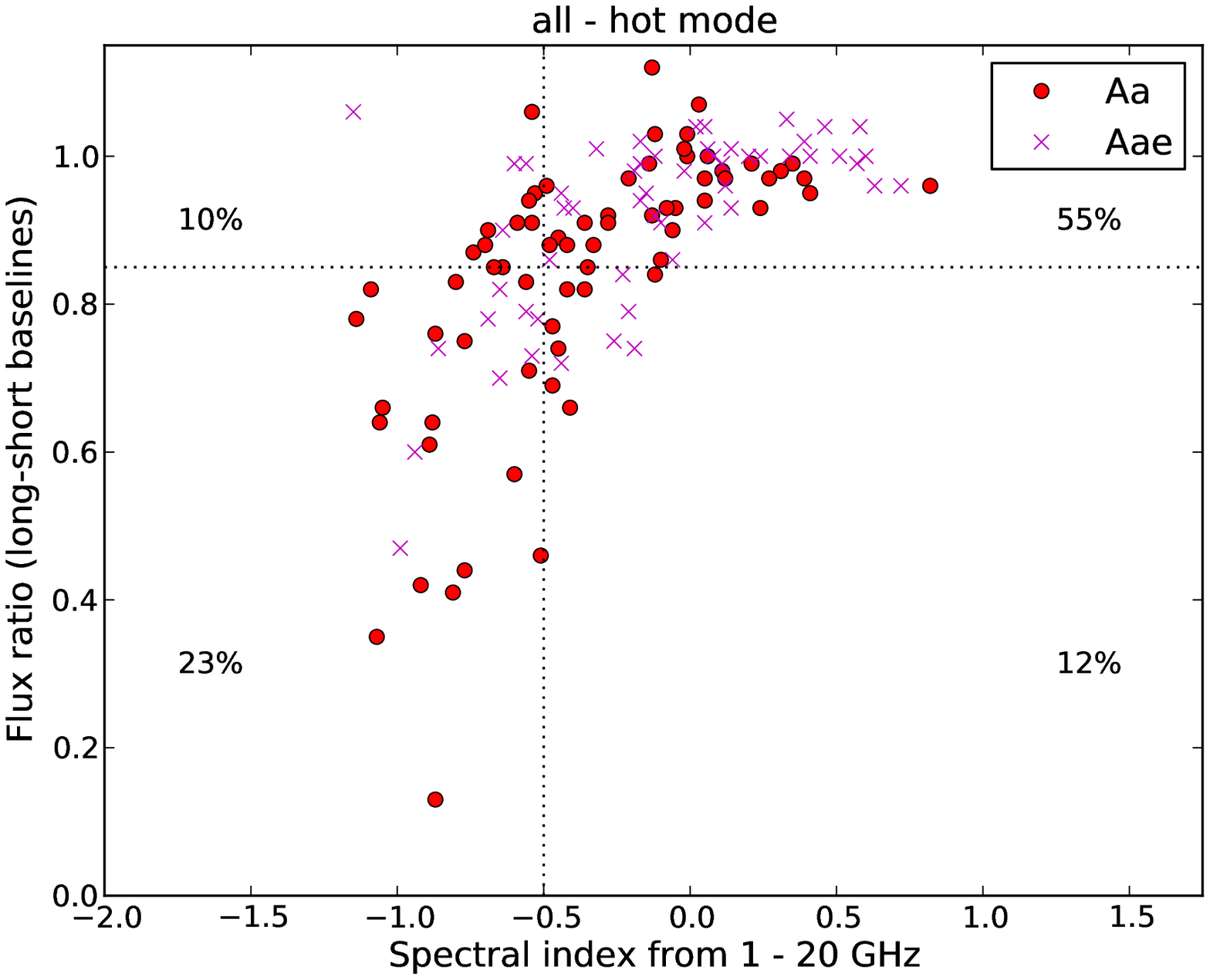, width=\linewidth}
\end{minipage}
\hspace{-0.5cm}
\begin{minipage}{0.5\linewidth}
\epsfig{file=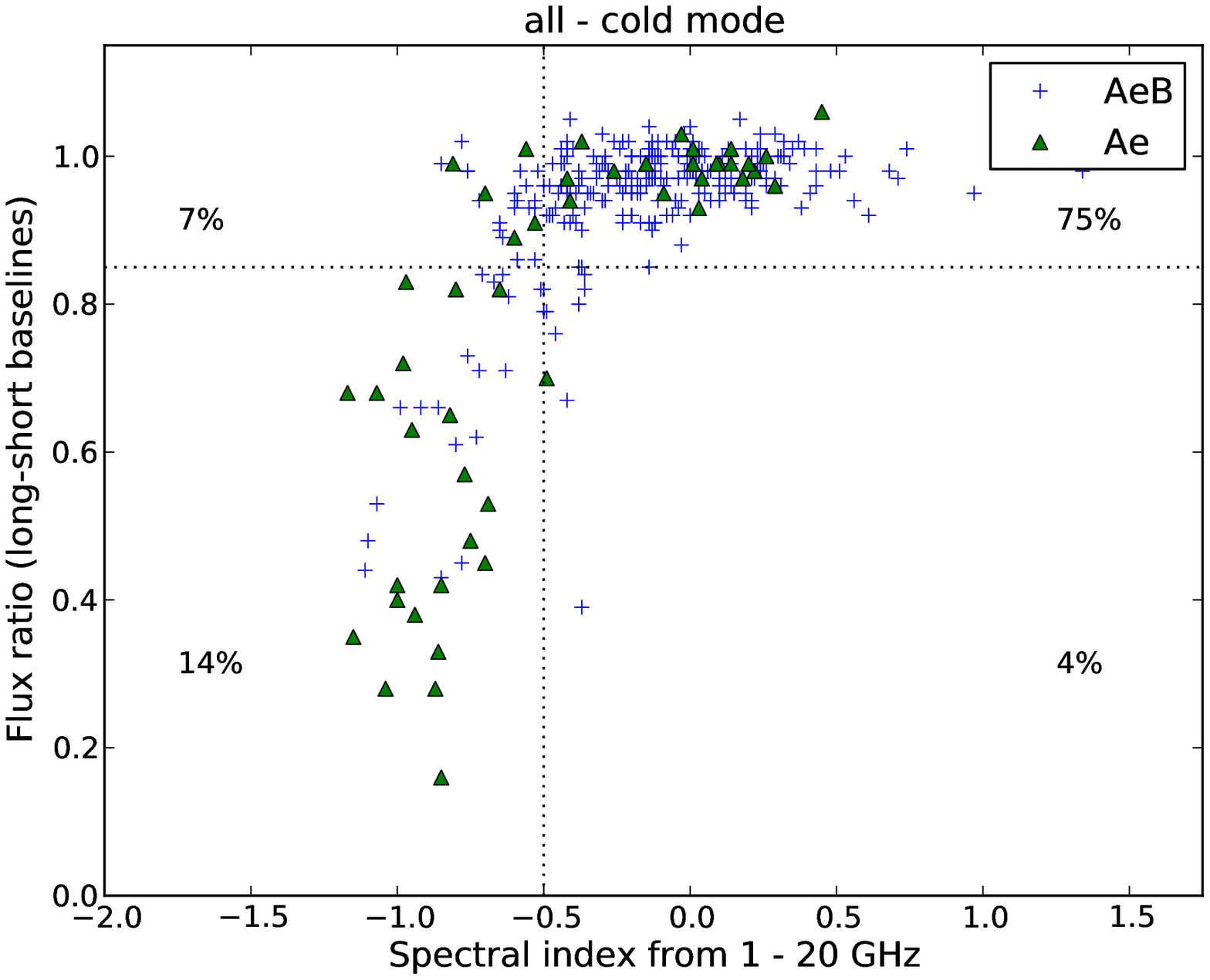, width=\linewidth}
\end{minipage}
\begin{minipage}{0.5\linewidth}
\epsfig{file=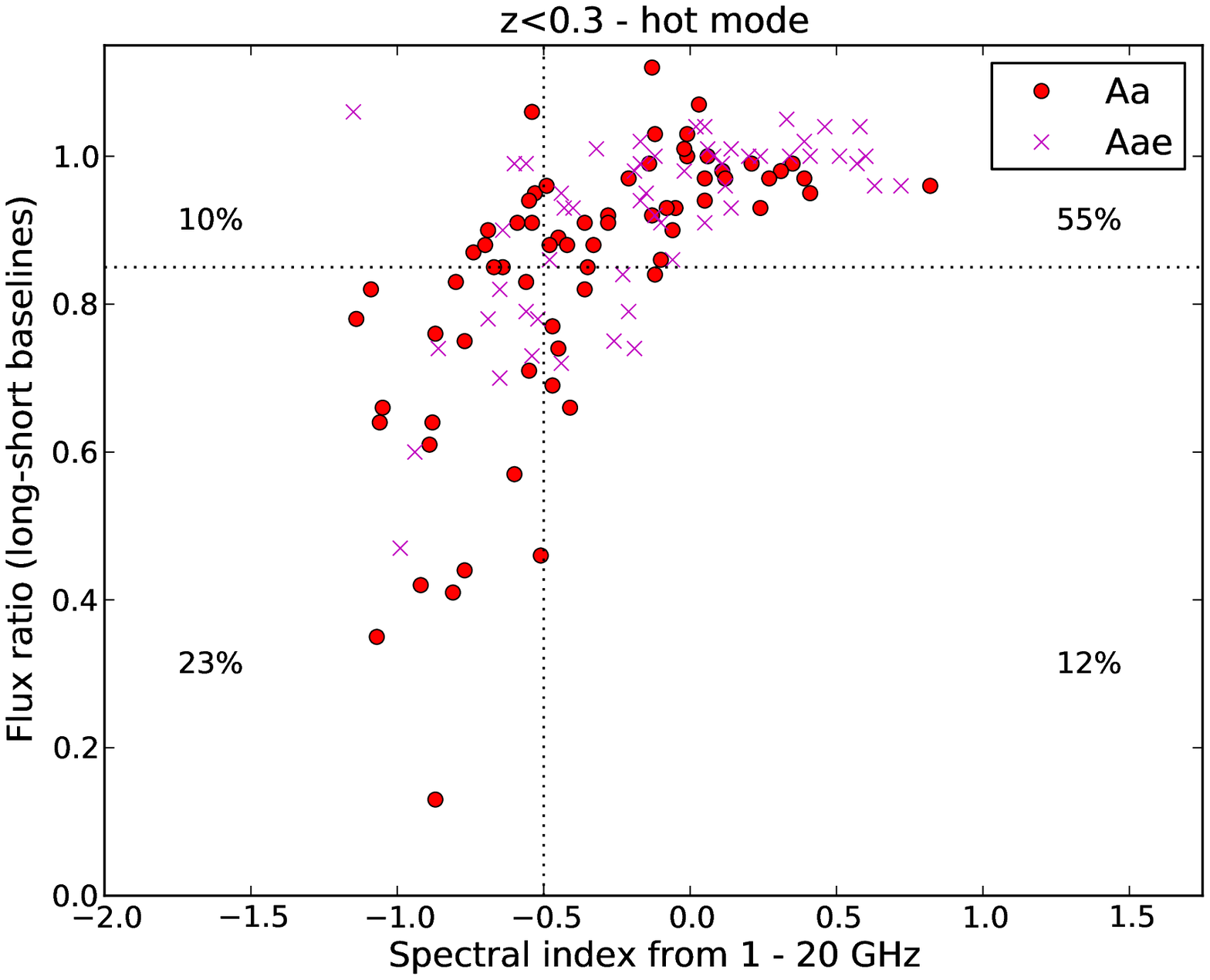, width=\linewidth}
\end{minipage}
\hspace{-0.5cm}
\begin{minipage}{0.5\linewidth}
\epsfig{file=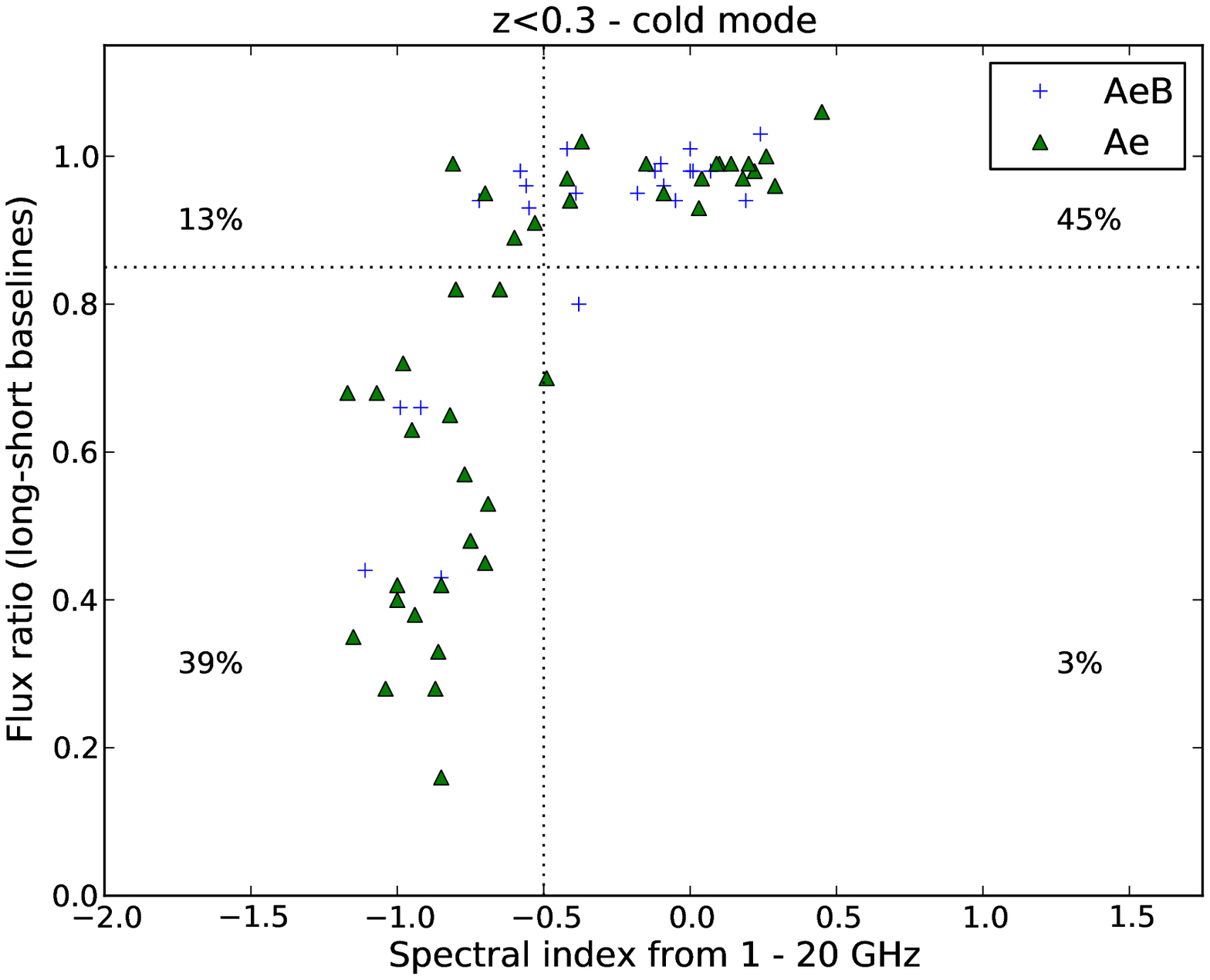, width=\linewidth}
\end{minipage}
\caption{Ratio of flux observed on the long--short baselines against spectral index. The hot and cold-mode sources have been plotted separately for the full spectroscopic sample (top) and for those sources with redshifts less than 0.3 (bottom). The dashed lines show the division between the compact and extended radio populations (flux ratio = 0.86) and flat and steep-spectrum populations ($\alpha=-0.5$). The percentages shown indicate the percentage of sources that fall into each quadrant. \label{accmode} }
\end{figure*}

Figure~\ref{6kvis} can be divided into 4 quadrants based on the observed radio properties; compact and flat-spectrum (upper right), compact and steep-spectrum (upper left), extended and steep-spectrum (lower left) and extended and flat-spectrum (lower right). It is not immediately obvious that there are any differences in the distribution of hot and cold-mode sources in this figure. For clarity, Figure~\ref{accmode} shows the same data with the hot and cold-mode objects separated (top panel), along with sources in the same redshift range ($z<0.3$) to eliminate any redshift dependencies (bottom panel). We have also indicated the percentage of objects that fall into each quadrant. At redshifts higher than $z=0.3$ the number of objects undergoing hot-mode accretion is too small to compare with the cold-mode sources. As indicated by the percentages shown in Figure~\ref{accmode}, the cold-mode sources above this redshift are virtually all compact, predominately due to the observational effect that at higher redshifts the linear sizes of the radio sources must be much larger to be a similar angular size as observed at lower redshifts.

It should be noted that at a redshift of $z=0.3$ the division between compact and extended sources ($\sim0.15$\,arcsec) corresponds to a linear size of 0.66\,kpc, therefore separating the core and inner jet from the large scale radio lobes. The differences between the two different accretion modes become more pronounced when comparing sources within the same redshift range. While both accretion modes can result in a range of radio properties, objects which are accreting in the cold-mode have a higher fraction of steep-spectrum, extended sources. This can also be seen in Table~\ref{flatsteepclassz03} where cold-mode sources have a steeper median spectral index. This once again suggests that objects that are accreting more efficiently have a greater chance of producing more luminous radio jets and lobes.  
 
\subsection{Optical vs. Radio luminosities}

\begin{figure}
\epsfig{file=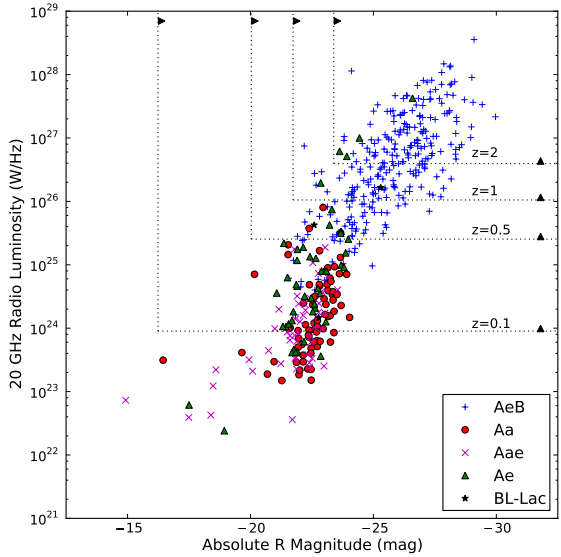, width=\linewidth}
\caption{20\,GHz radio luminosity against absolute B magnitude for sources in the AT20G spectroscopic sample. The survey limits of $S_{20}=40$\,mJy in the radio and R $=22$ in the optical are shown by the dashed lines for different redshifts. The sample has been plotted on equal axes; i.e. 5 magnitudes in the optical corresponds to a factor of 100 in radio luminosity. \label{absmagrlum}}
\end{figure}

The relationship between the rest frame 20\,GHz radio luminosity and R-band absolute magnitude for each spectral class in the spectroscopic sample is shown in Figure~\ref{absmagrlum}. The dashed lines denote the survey limits ($S_{20}=40$\,mJy and R $=22$) at different redshifts. 

Figure~\ref{absmagrlum} shows a clear distinction between the two main AT20G source populations; the higher luminosity quasar population (denoted by the blue `+' AeB sources) and the lower luminosity radio galaxies that reside in nearby massive ellipticals (mainly red `o' and magenta `x' points, corresponding to the Aa and Aae objects). This is largely a redshift effect with quasars observed out to higher redshifts due to their higher luminosities whilst the lower luminosity sources are only seen at lower redshifts before they fall out of the sample due to the flux limit of the survey. The majority of sources (both hot and cold-mode) also have absolute magnitudes brighter than $M_{\rm R}\sim-22$, consistent with the fact that AGN preferentially reside in more massive galaxies \citep{ledlow+owen, mauch07}.

\begin{figure}
\epsfig{file=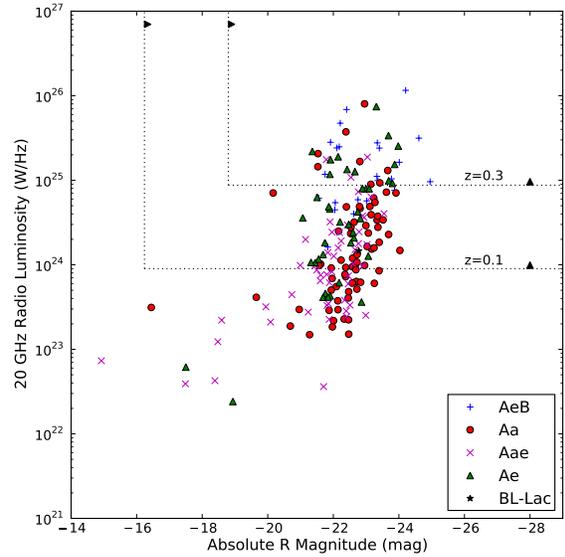, width=\linewidth}
\caption{Absolute R band magnitude against 20\,GHz radio luminosity for sources with redshifts $z<0.3$. \label{absmagrlumz03}}
\end{figure}

To study any differences between the hot and cold-mode populations we have again focused on those sources with $z<0.3$ as shown in Figure~\ref{absmagrlumz03}. Although there is no clear-cut separation, it can be seen that sources accreting in the hot-mode (Aa and Aae) have slightly higher absolute magnitudes for a given radio luminosity. That is, the red circles (Aa sources) have a tendancy to be to the right of the green triangles (Ae sources). Using the absolute R magnitude as a proxy for the stellar mass of the galaxy (i.e. more massive galaxies are likely to be brighter in the R band), this suggests that hot-mode accreting sources generally live in more massive host galaxies than the emission line, or cold-mode accreting sources, in agreement with the results of \citet{best05}. In addition, it has been shown that cold-mode sources tend to have younger stellar populations (\citealt{kauffmann08}, Ching et al. 2011, in preparation) which would further enhance the difference in absolute magnitude. 

Figure~\ref{absmagrlumz03} also shows that for a given absolute magnitude, the cold-mode sources have higher radio luminosities. Beaming effects are likely to contribute to the enhanced luminosities seen in AeB sources, but there are still more Ae sources (which would not be subject to beaming effects) at higher luminosities than Aa or Aae sources. Hence, cold-mode accreting sources have (on average) higher radio luminosities, but live in smaller host galaxies. As the size of the central SMBH is correlated with the size of the host galaxy \citep{magorrian, msigma}, this provides further evidence that it is the accretion rate, not the mass of the SMBH, that relates to the radio power of the jets and lobes \citep{2006MNRAS.372...21A, 2008A&A...486..119B, 2011MNRAS.411.1909F}.

Although the separation of hot and cold-mode accreting sources is by no means distinct, it appears to be loosely based on both the radio luminosity and absolute magnitude, remarkably similar to the `Ledlow \& Owen' division of FRI and FRII radio sources \citep{ledlow+owen}. However, this is most likely a side-effect of the fact that both cold-mode sources and FRII radio galaxies tend to have higher radio luminosities rather than evidence that the accretion mode is responsible for the FRI/FRII dichotomy (see also \citealt{best09}). 

Studying the optical properties of the 20\,GHz population provides insight into the accretion mechanism of these sources, complementing the radio data. We find that sources which are accreting efficiently have more powerful radio jets. However, the current spectroscopic dataset is subject to many selection effects and biases, and a complete sample is essential in investigating these correlations further.

\section{Comparison with S-cubed 18\,GHz selected sample} \label{skads}

As the AT20G survey is the largest, blind survey at 20\,GHz, it provides an important legacy catalogue for studies of galaxy and AGN evolution in the SKA era. To test the current predictions of what will be observed with the SKA, we have compared the AT20G catalogue with the S-cubed semi-emperical simulation (S3-SEX); a simulation of radio continuum sources out to $z=20$ \citep{s3-sex}. The resulting database contains flux measurements at 151\,MHz, 610\,MHz, 1.4\,GHz, 4.86\,GHz and 18\,GHz, down to the flux limit of 10\,nJy in an area of $20\times20$\,deg$^{2}$.

\begin{figure}
\epsfig{file=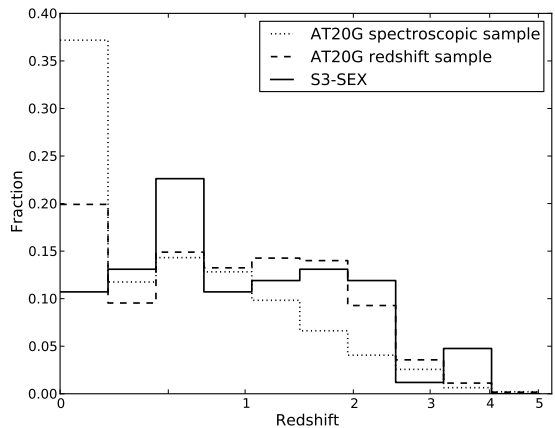, width=\linewidth}
\caption{Redshift distributions (binned in log($1+z$)) of the AT20G redshift sample (dashed line) and spectroscopic sample (dotted line) compared to the 18\,GHz selected S3-SEX simulated dataset (solid line). Due to the large difference in the number of sources in each sample, the y-axis shows the fraction of sources that fall in each bin. \label{skadsz}}
\end{figure}

For comparison with the AT20G survey, we have selected sources from the S3-SEX database that have an 18\,GHz flux measurement above 100\,mJy (the completeness limit of AT20G). Figure~\ref{skadsz} shows the redshift distributions of the S3-SEX dataset and the AT20G survey, normalised due to the different sample sizes. The AT20G redshift sample is roughly similar to the 18\,GHz S3-SEX sample and K-S tests suggest they are drawn from the same population. However, the AT20G spectroscopic sample (shown be the dotted line) is very different to the S3-SEX distribution and suggests that we are missing a lot of high-redshift sources. This is unsuprising as most of the AT20G spectroscopic sample were observed as part of the 6dF Galaxy Survey and is therefore biased towards sources at lower redshifts.

\begin{figure}
\epsfig{file=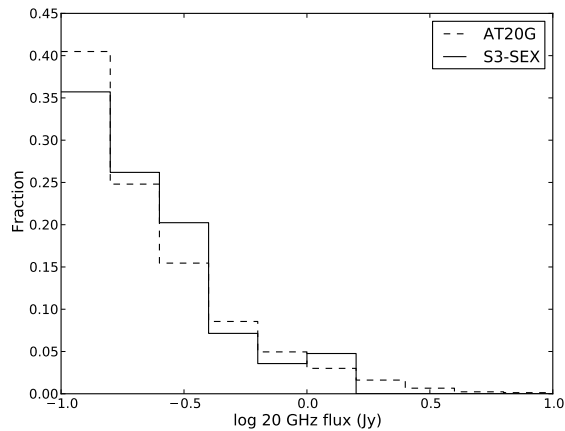, width=\linewidth}
\caption{Normalised flux distributions of the AT20G survey (dashed line) and the 18\,GHz selected S3-SEX dataset (solid line). \label{skadsflux}}
\end{figure}

\begin{figure}
\epsfig{file=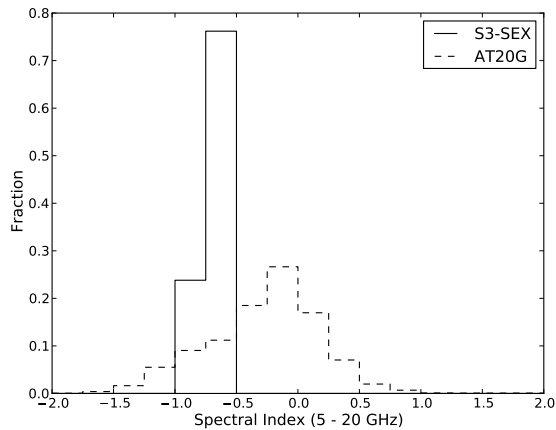, width=\linewidth}
\caption{Spectral index distributions of the AT20G survey (dashed line) and the 18\,GHz selected S3-SEX catalogue (solid line). \label{skadsalpha}}
\end{figure}

To ensure that the 18\,GHz selected S3-SEX sample is comparable to the AT20G survey we have also compared the radio properties of the two samples. Figure~\ref{skadsflux} shows the flux distribution of each sample and a K-S test carried out on these distributions indicate they are drawn from the same parent population. However, the spectral indices (from 5--20\,GHz) of the S3-SEX selected sample are significantly different from AT20G as shown in Figure~\ref{skadsalpha}. Sources selected at high radio frequencies are predominately flat-spectrum due to the dominance of the compact core. However, selecting sources at 18\,GHz from the S3-SEX database produces a catalogue entirely comprised of steep-spectrum sources. 

This is an artifact of the input models used to generate properties of radio-loud AGN in the semi-empirical simulation. As described in \citet{s3-sex} the dataset is drawn from a 151\,MHz population and extrapolated to higher frequencies assuming a power law ($\alpha=-0.75$) for the extended emission, while the core emission was modelled based on the results of \citet{jarvisrawlings}, who studied a sample of flat-spectrum quasars selected at 2.7\,GHz. The population of compact sources which dominate high-frequency surveys are hardly represented in a 151\,MHz selected catalogue, therefore extrapolating sources from low frequencies will not correctly predict the source population at high frequencies. 

It was noted by \citet{s3-sex} that the S3-SEX models cannot simply be extrapolated to higher frequencies. They also noted that the simulation underestimates the number of flat-spectrum sources observed, leading to discrepancies between the simulated and observed spectral index distributions, even at lower frequencies. This discrepancy only gets compounded as we move to higher frequencies as the flat-spectrum sources become the dominant population. 

This highlights the need for large area, high-frequency data which is provided by the AT20G survey. Studying both the optical and radio properties of the high-frequency population provides much needed information on the compact cores of the AGN and is essential in understanding AGN and galaxy evolution. The AT20G catalogue provides a valuable database that can be used to update the S3-SEX models as well as any future SKA-related simulations.

\section{Summary} \label{conclusions}

The primary goal of this paper is to present a catalogue of the optical identifications and redshifts for the Australia Telescope 20\,GHz (AT20G) survey. The positional accuracy ($\sim 1$\,arcsec) of AT20G sources, and the dominance of nuclear point sources meant that the majority of sources were identified using an automated process. Optical identifications were made using the SuperCOSMOS Science Archive with 78.5\% of AT20G sources (with $|b|>10^{\circ}$) having optical counterparts, much larger than is seen at lower radio frequencies (typically 25--30\%, \citealt{bock99}). The optical identification rate increases as a function of 20\,GHz flux density, predominately due to the increase in `stellar' optical IDs with increasing flux density. On the other hand, sources with `galaxy' counterparts (as determined by SuperCOSMOS based on the optical morphology) decreased with increasing flux density. Whilst the `stellar' sources are the dominant population in the AT20G survey, we predict that if we were to probe deeper at 20\,GHz the galaxy population would become dominant. 

Redshifts were found from either the 6dF Galaxy Survey \citep{6df} or from the literature using NED. We also present 144 new redshifts that were obtained from poor-weather backup programs from a range of facilities. A total of 30.9\% of the AT20G survey have redshift information. The redshift completeness is quite high for the bright AT20G sources (91.5\% complete for sources brighter than 1\,Jy and 85.3\% for sources above 500\,mJy), but this drops off rapidly towards fainter sources.  

Objects that have an optical spectrum available (i.e. from either the 6dF survey or additional observations) formed the `AT20G spectroscopic sample' and this sample is used to study the emission line properties of the sample. This is achieved by dividing the sample into `hot' or `cold' mode accretors, where sources with strong emission lines are classified as undergoing cold-mode accretion and sources with weak or no emission lines present were classified as hot-mode. This naming convention arises from the fact that cold-mode sources primarily accrete cold gas, forming a radiatively efficient accretion disk which fits the conventional AGN unification models. On the other hand, hot-mode sources accrete hot gas which leads to an inefficient accretion disk and none of the observational criteria normally associated with optical or X-ray selected AGN.

It is found that cold-mode accreting sources have a steeper median radio spectral index, suggesting that the radio jets of these sources dominate the emission (rather than the cores), consistent with the idea that these sources are more efficient accretors. It is also found that the cold-mode accretors show different orientation effects; narrow emission line objects (Ae sources) have a larger fraction of steep-spectrum radio sources, while objects with broad emission lines (AeB sources) are dominated by flat-spectrum sources as predicted by unification models \citep{antonucci93, urry+padovani}. On the other hand, the hot-mode accretors show no distinct orientation effects.

Due to the high-frequency selection, the AT20G sample is dominated by flat-spectrum sources, as are many of the subsamples. AT20G sources with optical IDs, the redshift sample and the spectroscopic sample are all dominated by flat-spectrum sources, yet for the subset of sources that do not have an observed optical counterpart (blank fields) the split into flat and steep spectrum sources is approximately 50/50. This implies that many of these blank fields exhibit narrow emission lines (falling into the Ae spectral class) as this is the only population of sources dominated by steep-spectrum sources. Some of these blank fields also form a small sample (36 objects) of high-frequency ultra-steep spectrum (USS) sources, listed in Appendix \ref{appendixb}. As the peak of the radio SED is thought to move down in frequency as the source ages, it is likely that these are the progenitors of the USS sources selected at lower frequencies.

Studying the radio luminosity vs. absolute magnitude distribution reveals that, for a given radio luminosity, the cold-mode accretors are more likely to reside in smaller host galaxies whilst the hot-mode accreting sources generally live in more massive galaxies. Objects accreting in the cold-mode also have higher radio luminosities for the same absolute magnitude than their hot-mode counterparts, providing further evidence that the cold-mode accreting sources produce more luminous radio jets and lobes. 

Lastly, we compared the AT20G survey with the S-cubed semi-empirical simulations (S3-SEX). This was achieved by selecting sources from the S3-SEX database to be brighter than 100\,mJy (the AT20G completeness limit) at 18\,GHz. The flux distributions of the S3-SEX and AT20G samples are in agreement with each other, but the spectral index distributions are remarkably different, implying that the models used to determine the S3-SEX core fluxes need refining. This comparison highlights the need for large samples of high-frequency radio sources provided by the AT20G survey.

We continue to compile spectroscopic information on AT20G sources with the aim of forming a complete sample to further study the properties of high-frequency radio sources and how this population evolves over cosmic time.

\section*{Acknowledgments}

EKM would like to thank Garret Cotter, Angela Taylor, Mike Jones, Tom Mauch and Steve Rawlings for many useful discussions concerning the S3-SEX database, as well as for the funding provided while visiting Oxford in October, 2010. We also thank the referee for valuable comments which helped improve the paper. We acknowledge the support of the Australian Research Council through the award of an ARC Australian Professorial Fellowship (DP0451395) to EMS, an ARC QEII Fellowship (DP0666615) to SMC, a Federation Fellowship (FF0345330) to RDE and an ARC Australian Postdoctoral Fellowship (DP0665973) to TM. MM acknowledges partial financial support by ASI (ASI/INAF Agreement I/072/09/0 for the Planck LFI activity of Phase E2 and contract I/016/07/0 COFIS).

This research has made use of data obtained from the SuperCOSMOS Science Archive and 6dF Galaxy Survey database, prepared and hosted by the Wide Field Astronomy Unit, Institute for Astronomy, University of Edinburgh, which is funded by the UK Science and Technology Facilities Council. This research has also made use of the NASA/IPAC Extragalactic Database (NED) which is operated by the Jet Propulsion Laboratory, California Institute of Technology, under contract with the National Aeronautics and Space Administration. The authors have also presented data based on observations obtained at the Gemini Observatory (program GS-2009B-Q-95), which is operated by the Association of Universities for Research in Astronomy, Inc., under a cooperative agreement with the NSF on behalf of the Gemini partnership: the National Science Foundation (United States), the Science and Technology Facilities Council (United Kingdom), the National Research Council (Canada), CONICYT (Chile), the Australian Research Council (Australia), Minist\'{e}rio da Ci\^{e}ncia e Tecnologia (Brazil) and Ministerio de Ciencia, Tecnolog\'{i}a e Innovaci\'{o}n Productiva (Argentina).

\bibliographystyle{scemnras}

\setlength{\bibhang}{2.0em}
\setlength\labelwidth{0.0em}
\bibliography{bibliog}
  
\appendix
\section[Appendix A]{Gemini spectra} \label{appendixa}
\begin{figure*}
\epsfig{file=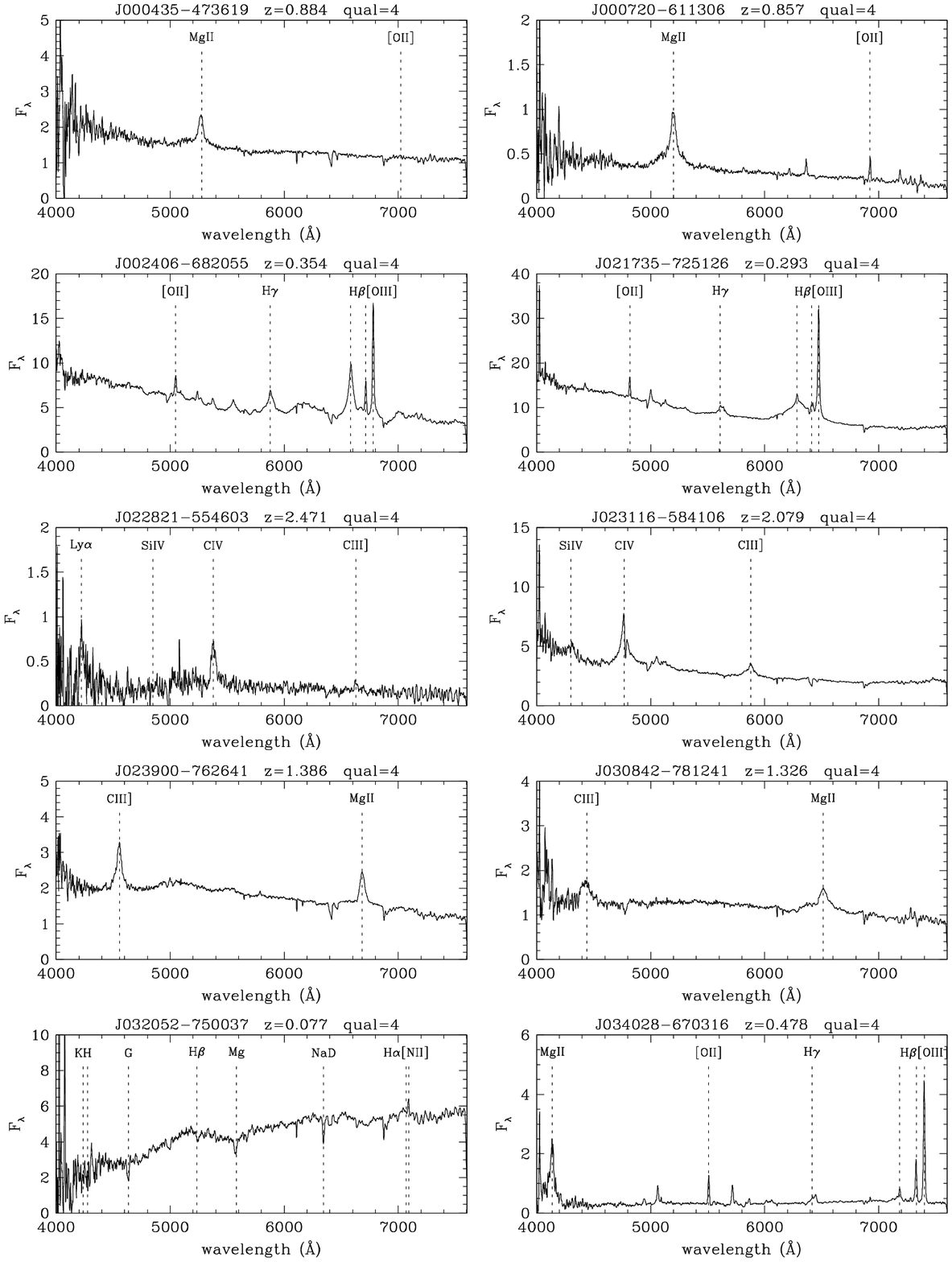, width=\linewidth}
\caption{Spectra obtained through the Gemini poor weather program (program ID: GS-2009B-Q-95). The spectra have been smoothed using a boxcar algorithm ver 5 pixels, corresponding to $\sim 7 {\rm{\AA}}$.}
\end{figure*} 
\begin{figure*}
\epsfig{file=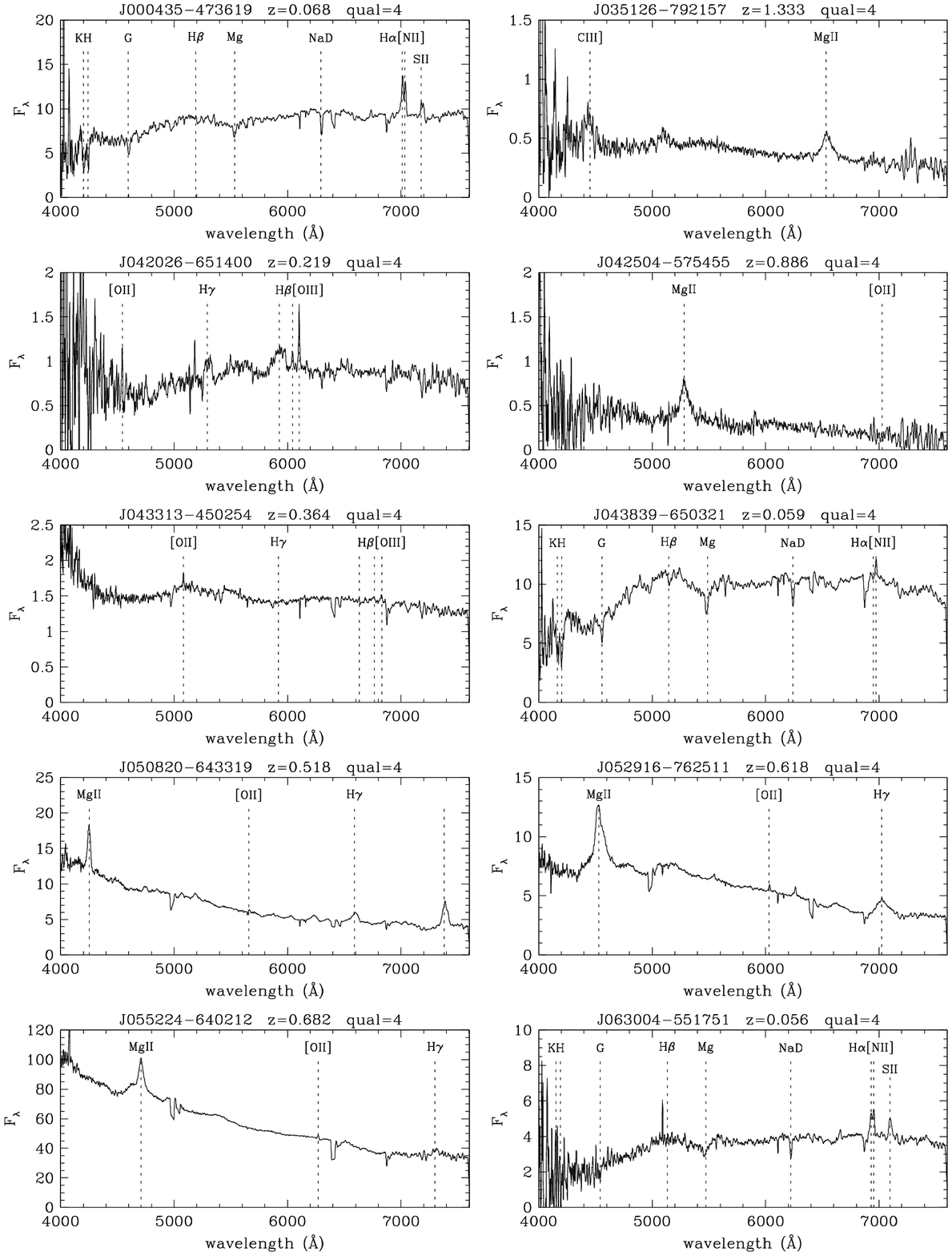, width=\linewidth}
\caption{Spectra obtained through the Gemini poor weather program (program ID: GS-2009B-Q-95).}
\end{figure*} 
\begin{figure*}
\epsfig{file=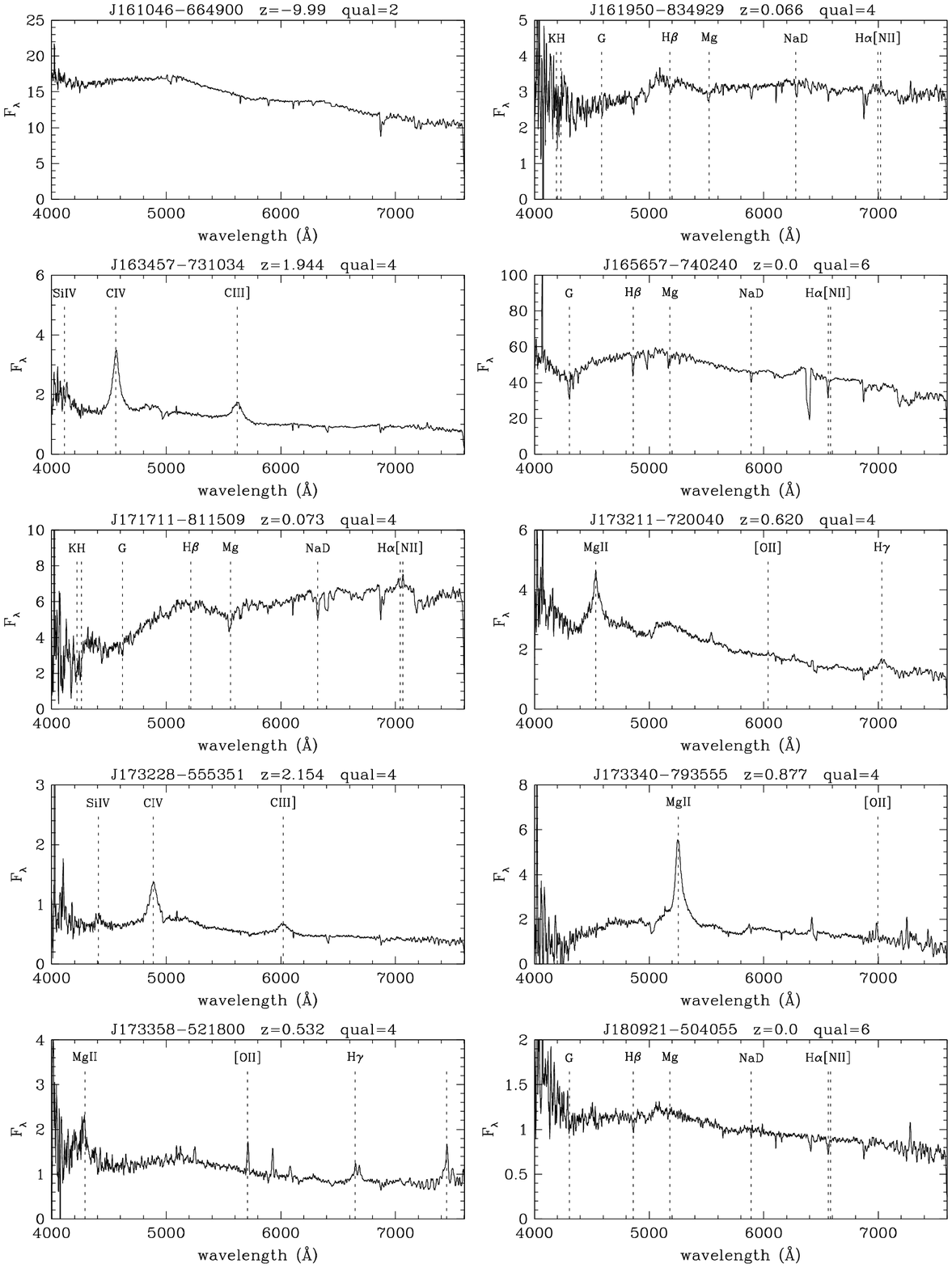, width=\linewidth}
\caption{Spectra obtained through the Gemini poor weather program (program ID: GS-2009B-Q-95).}
\end{figure*} 
\begin{figure*}
\epsfig{file=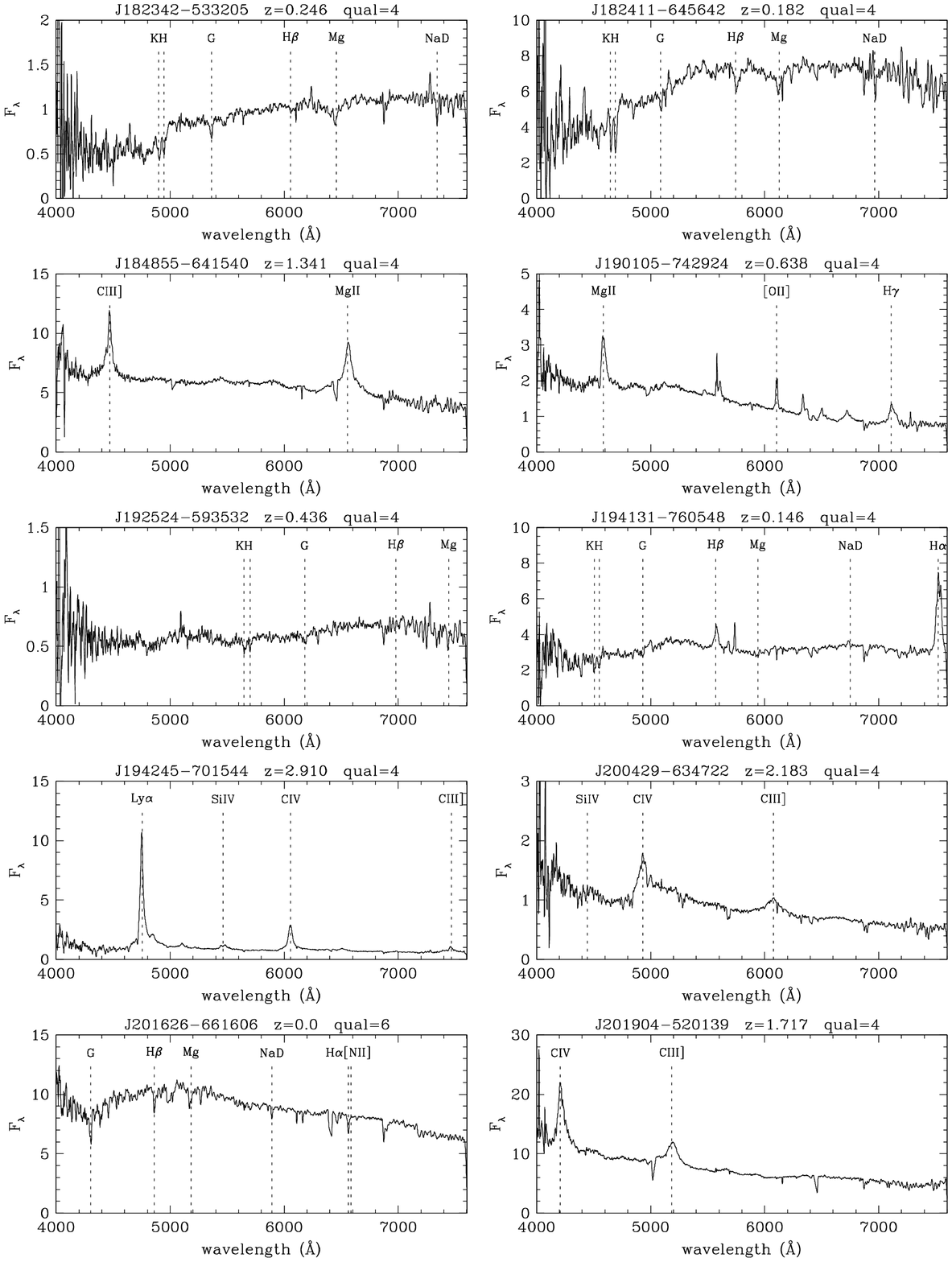, width=\linewidth}
\caption{Spectra obtained through the Gemini poor weather program (program ID: GS-2009B-Q-95).}
\end{figure*} 
\begin{figure*}
\epsfig{file=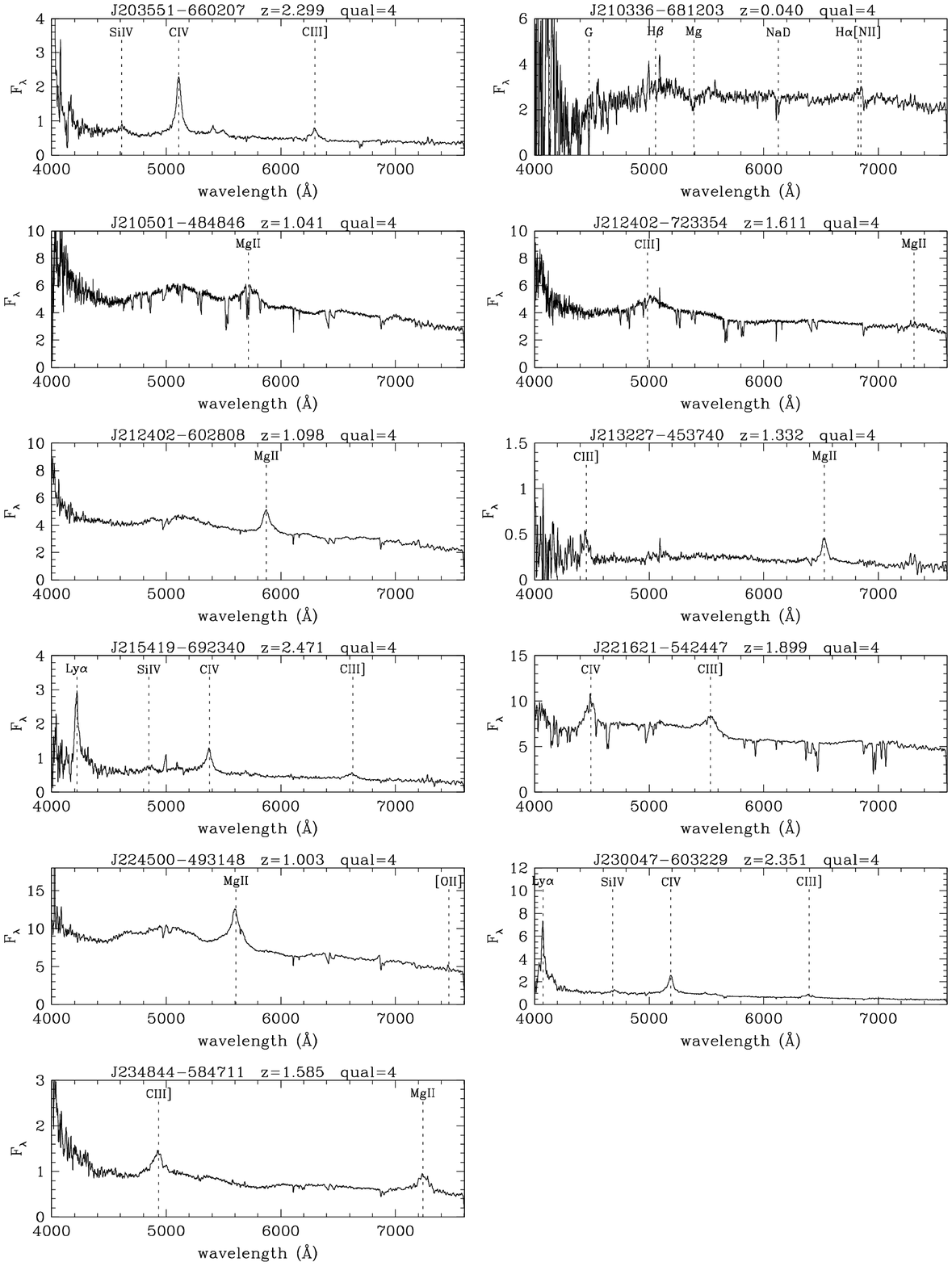, width=\linewidth}
\caption{Spectra obtained through the Gemini poor weather program (program ID: GS-2009B-Q-95).}
\end{figure*} 

\section[Appendix B]{USS sources} \label{appendixb}

\begin{table*}
\begin{center}
\caption{List of sources with an ultra-steep spectrum ($\alpha_5^{20} <-1.2$) and no optical counterpart as discussed in Section~\ref{flatvssteep}. The typical error on the spectral index is 0.06 \citep{at20ganalysis}. Redshifts were obtained from NED. \label{uss}}
\begin{tabular}{lccccc}
\hline
{\bf AT20G name} & {\bf $\alpha_5^{20}$} & {\bf $\alpha_1^{5}$} & {\bf Other name} & {\bf Redshift} & {\bf Reference} \\ 
\hline
AT20GJ001052$-$415310 & $-$1.53 & $-$0.87 & PKS 0008$-$42 & 1.12 (PHOT) & \citet{ms4} \\
AT20GJ001601$-$631009 & $-$1.26 & $-$1.11 & PKS 0013$-$63 & --- & --- \\
AT20GJ010837$-$285128 & $-$1.45 & $-$0.81 & PKS 0106$-$291 & --- & --- \\ 
AT20GJ011539$-$281712 & $-$1.33 & $-$0.64 & PKS 0113$-$285 & --- & --- \\
AT20GJ011651$-$205206 & $-$1.36 & $-$0.73 & PKS 0114$-$21 & 1.410 & \citet{1996ApJS..107...19M}\\
AT20GJ013115$-$510136 & $-$1.37 & $-$0.97 & PKS 0129$-$51 & --- & --- \\
AT20GJ014127$-$270606 & $-$1.41 & $-$1.13 & PKS 0139$-$273 & 1.440 & \citet{1996ApJS..107...19M}\\
AT20GJ015201$-$294102 & $-$1.28 & $-$1.01 & PKS 0149$-$29 & 0.603 & \citet{1996ApJS..107...19M}\\
AT20GJ015952$-$743056 & $-$1.27 & $-$0.98 & PKS 0159$-$747 & --- & --- \\
AT20GJ025515$-$665655 & $-$1.26 & $-$0.86 & --- & --- & --- \\
AT20GJ031610$-$682105 & $-$1.34 & $-$1.05 & PKS 0315$-$68 & 1.30 (PHOT) & \citet{ms4} \\
AT20GJ041300$-$343009 & $-$1.21 & $-$1.00 & PKS 0411$-$34  & 0.83 (PHOT) & \citet{ms4} \\
AT20GJ042056$-$622338 & $-$1.25 & $-$0.97 & PKS 0420$-$62 & 0.81 (PHOT) & \citet{ms4} \\
AT20GJ044737$-$220336 & $-$1.34 & $-$1.03 & PKS 0445$-$22 & --- & --- \\
AT20GJ044832$-$524108 & $-$1.29 & $-$0.93 & PKS 0447$-$527 & --- & --- \\
AT20GJ054617$-$172552 & $-$1.23 & $-$0.78 & PKS 0544$-$17 &  --- & --- \\
AT20GJ064848$-$473426 & $-$1.35 & $-$0.86 & PKS 0647$-$475 & 1.94 (PHOT) & \citet{ms4} \\
AT20GJ072232$-$370412 & $-$1.25 & $-$0.77 & PMN J0722$-$3704 & --- & --- \\
AT20GJ084600$-$261054 & $-$1.36 & $-$0.77 & PKS 0843$-$260 & --- & --- \\
AT20GJ085300$-$204721 & $-$1.47 & $-$1.05 & PKS 0850$-$20 & 1.337 & \citet{1999MNRAS.310..223B} \\
AT20GJ090652$-$682939 & $-$1.23 & $-$1.01 & PKS 0906$-$682 & --- & --- \\
AT20GJ103828$-$700308 & $-$1.26 & $-$0.94 & PKS 1036$-$69 & 0.87 (PHOT) & \citet{ms4} \\
AT20GJ105132$-$202346 & $-$1.40 & $-$0.96 & PKS 1049$-$201 & 1.116 & \citet{1996ApJS..107...19M} \\
AT20GJ110622$-$210858 & $-$1.33 & $-$0.99 & PKS 1103$-$20 & 1.120 & \citet{1996ApJS..107...19M} \\
AT20GJ131829$-$462035 & $-$1.67 & $-$0.84 & PMN J1318$-$4620 & 1.12 (PHOT) & \citet{ms4} \\
AT20GJ145104$-$750935 & $-$1.56 & $-$0.87 & PKS 1445$-$74 & --- & --- \\
AT20GJ161933$-$841819 & $-$1.33 & $-$0.97 & PKS 1607$-$84 & 1.11 (PHOT) & \citet{ms4} \\
AT20GJ172516$-$800428 & $-$1.41 & $-$0.50 & --- & --- & --- \\
AT20GJ172649$-$552939 & $-$1.58 & $-$0.47 & PMN J1726$-$5529 & --- & --- \\
AT20GJ172657$-$642753 & $-$1.41 & $-$0.64 & PMN J1726$-$6427 & --- & --- \\
AT20GJ175906$-$594702 & $-$1.60 & $-$1.16 & PKS 1754$-$59  & 0.80 (PHOT) & \citet{ms4} \\
AT20GJ185841$-$631306 & $-$1.24 & $-$0.97 & PKS 1853$-$632 & --- & --- \\
AT20GJ195715$-$422220 & $-$1.28 & $-$0.98 & PKS 1953$-$42 & 0.82 (PHOT) & \citet{ms4} \\
AT20GJ232207$-$544528 & $-$1.21 & $-$1.03 & PKS 2319$-$55 & 0.73 (PHOT) & \citet{ms4} \\
AT20GJ232634$-$402717 & $-$1.25 & $-$0.88 & PKS 2323$-$40 & 0.81 (PHOT) & \citet{ms4} \\
AT20GJ233512$-$663709 & $-$1.22 & $-$0.65 & PKS 2332$-$66 &  --- & --- \\
\hline
\end{tabular}
\end{center}
\end{table*}

\bsp
  
\label{lastpage}

\end{document}